\documentclass[12pt,reqno]{amsart}

\usepackage[T1]{fontenc}
\usepackage[utf8]{inputenc}
\usepackage[english]{babel}

\usepackage{a4wide, mathrsfs, csquotes, amsfonts, amssymb, amsmath, mathtools, enumerate, xcolor, dsfont, mathabx}

\usepackage{graphicx}
\usepackage[font=small,labelfont=bf]{caption}
\usepackage[subrefformat=parens]{subcaption}

\usepackage{pgf, tikz, pgfplots}
\pgfplotsset{compat=1.16}
\usetikzlibrary{arrows}

\theoremstyle{plain}
\newtheorem{theorem}{Theorem}[section]
\newtheorem{lemma}[theorem]{Lemma}

\newtheorem{proposition}[theorem]{Proposition}

\theoremstyle{definition}
\newtheorem{remark}[theorem]{Remark}
\newtheorem{notation}{Notation}

\definecolor{bleusombre}{rgb}{0,0,0.6}
\definecolor{Bl}{rgb}{0,0,0.6}
\definecolor{rougesombre}{rgb}{0.8,0,0}
\definecolor{vertsombre}{rgb}{0,0.6,0}
\definecolor{webblue}{rgb}{0.22,0.45,0.70}
\definecolor{webred}{rgb}{0.5, 0.09, 0.09}
\definecolor{zzttqq}{rgb}{0.6,0.2,0.}

\usepackage[plainpages=false,colorlinks,linkcolor=bleusombre,
citecolor=rougesombre,urlcolor=vertsombre,breaklinks]{hyperref}

\newcommand{\Z}{\mathbb{Z}}

\renewcommand{\leq}{\leqslant}	
\renewcommand{\geq}{\geqslant}
\newcommand{\C}{\mathbb{C}}
\newcommand{\R}{\mathbb{R}}

\newcommand{\1}{{\bf 1}}
\newcommand{\sign}{{\mathrm{sign}\,}}

\makeatletter

\@addtoreset{equation}{section}
\makeatother

\numberwithin{equation}{section}

\title[Green's functions for Dirac operators]{Green's functions for magnetic Dirac operators and bulk-edge correspondence}
\author[J.-M. Barbaroux]{J.-M. Barbaroux}
\address[J.-M. Barbaroux]{Univ Toulon, Aix Marseille Univ, CNRS, CPT, Toulon, France}
\email{barbarou@univ-tln.fr}

\author{H. D. Cornean}
\address[H. D. Cornean]{Department of Mathematical Sciences, Aalborg University, Skjernvej 4A, 9220 Aalborg, Denmark}
\email{cornean@math.aau.dk}

\author[L. Le Treust]{L. Le Treust}
\address[L. Le Treust]{Aix Marseille Univ, CNRS, I2M, Marseille, France}
\email{loic.le-treust@univ-amu.fr}

\author[N. Raymond]{N. Raymond}
\address[N. Raymond]{Univ Angers, CNRS, LAREMA, Institut Universitaire de France, F-49000 Angers, France}
\email{nicolas.raymond@univ-angers.fr }

\author[E. Stockmeyer]{E. Stockmeyer}
\address[E. Stockmeyer]{Instituto de F\'isica, Pontificia Universidad Cat\'olica de Chile, Vicu\~na Mackenna 4860, Santiago 7820436, Chile.}
\email{stock@fis.puc.cl}

\begin{document}
	
	\begin{abstract}
	 We study the resolvent kernel (Green's function) of magnetic Dirac operators on a half-plane with boundary conditions interpolating between infinite mass and zigzag cases, excluding  the latter. We show that these kernels have all the required properties so that the proof of bulk-edge correspondence from the infinite mass case shown recently can be repeated \emph{ad litteram}. The zigzag case exhibits qualitatively different behavior and will be addressed in future work. 
		\end{abstract}
    \maketitle

	\tableofcontents
\newpage
    \section{Motivation and main results}
In this paper, we continue our investigation of the bulk-edge correspondence for magnetic Dirac systems defined on the half-plane, subject to a one-parameter family of translation-invariant boundary conditions that interpolate between the infinite mass and zigzag cases. In a recent work \cite{MR4769230}, we established in this setting  the bulk-edge correspondence for the purely translation-invariant magnetic Dirac operator  in the direction parallel to the edge. In particular, we identified an anomaly in the zigzag limit, associated with the emergence of zero-energy modes and the breakdown of ellipticity. That analysis heavily relied on translation invariance and focused on the spectral properties of the corresponding energy dispersion curves. Namely, in this ``simple'' case, one can compute by brute force the right hand side of \eqref{hhh4} which gives the edge conductance and compare it with the Chern number of the corresponding bulk Dirac--Landau projection. This approach breaks down completely if the operator is no longer purely magnetic, and another method is thus needed. It turns out that in the infinite mass case, one can prove bulk-edge correspondence by using certain properties of exponential localization and regularity near the diagonal of the integral kernel of the resolvent \cite{Cornean_2023}. In the current paper, we show that these properties remain true for a much larger class of boundary conditions which interpolate between infinite mass and zigzag, but breaks down for the zigzag case.

Thus our primary object of study is the resolvent kernel (Green's function) of these Dirac operators. We analyze its analytic structure and decay behavior and use these properties to extend the bulk-edge correspondence to settings that include bounded Hermitian matrix-valued perturbations of the homogeneous magnetic system.

\subsection{Main results}
Our main result shows that, for all non-zigzag boundary conditions, the resolvent kernel satisfies the continuity and exponential localization estimates required to reproduce---almost verbatim---the proof of the bulk-edge correspondence established in \cite{Cornean_2023} for the infinite mass case \cite{MR3950662, barbaroux2019resolvent}. The zigzag case is excluded from our analysis due to its qualitatively different spectral features and will be addressed in future work.
\begin{center}
$ $
\end{center}
Let us first turn to the definition of the model by 
introducing some notation. The half-plane will be denoted by
\[\mathbb{R}^2_+=\{x=(x_1,x_2)\in\mathbb{R}^2 : x_2>0\}\,.\]
For a homogeneous magnetic field orthogonal to the plane and of strength $b\neq 0$, we associate to it a magnetic vector potential $\mathbf{A}=(A_1,A_2)$ such that $b=\partial_1 A_2-\partial_2 A_1$ and in the Landau gauge it is given by
\begin{equation}\label{eq.ax_2}
\mathbf{A}(x)=(-bx_2,0)\, .
\end{equation}
The standard Pauli matrices are given by 
\[
\sigma_1 = \begin{pmatrix} 0 & 1 \\ 1 & 0 \end{pmatrix}\,,\quad 
\sigma_2 = \begin{pmatrix} 0 & -i \\ i & 0 \end{pmatrix}\,,\quad \text{and} \quad 
\sigma_3 = \begin{pmatrix} 1 & 0 \\ 0 & -1 \end{pmatrix}\,.
\]
The present article is devoted to the study of the resolvent of Dirac operators acting in $L^2(\R^2_+,\mathbb{C}^2)$  as
	\begin{equation}\label{tigre9} \mathscr{D}_{b,W}:=\sigma\cdot(-i\nabla-\mathbf{A})+W\,,
    \end{equation}
    where $W=w_0 \1_2 +\sum_{j=1}^3 w_j\, \sigma_j \in { L^\infty}(\R^2_+, \C^{2\times 2})$ is a matrix-valued multiplication operator with $\{w_k\}_{k=0}^3$ real-valued { and bounded}. At the edge, given by \[\partial\mathbb{R}^2_+=\{x=(x_1,x_2)\in\mathbb{R}^2 : x_2=0\}\,, \]
we may impose the most general local boundary condition depending on a constant $\gamma\in\R\cup\{+ \infty\}$:
\begin{equation}\label{bound.cond}\begin{cases}
		\psi_2(x_1,0)=\gamma \psi_1(x_1,0)&\text{ if }\gamma\in\R\,,
		\\
		\psi_1(x_1,0) = 0&\text{ if }\gamma = +\infty\,.
\end{cases}\end{equation} 
The infinite mass boundary condition corresponds to cases $\gamma=\pm 1$ and cases $\gamma \in \{0,+\infty\}$ yield the so-called zigzag boundary conditions. In this work we focus on the study of all non-zigzag local boundary conditions which are translation invariant. In view of the symmetries of the system (see \cite[Remark 1.10]{MR4769230}) it is enough to consider simply a constant $\gamma\in (0,+\infty) $. In this case the domain of self-adjointness of the operators $\mathscr{D}_{b,W}$ is known to be contained in the first Sobolev space $H^1(\R^2_+,\C^2)$ (see \cite[Theorem 1.15]{barbaroux:hal-02889558}).   

\begin{notation}\label{stock3}
Given $A=\{A_{jk}\}_{1\leq j,k\leq 2}$ a $2\times 2$ matrix, we set
\begin{equation}\label{stock2}
|A|:=\sup_{1\leq j,k\leq 2} |A_{jk}|\, .
\end{equation}
Moreover, $A^\dagger$ is the matrix given by complex conjugation of the transpose of $A$: 
\begin{equation}\label{stock4}
(A^\dagger)_{jk}=\overline{A_{kj}}\, .
\end{equation}
\end{notation}

We are now ready to formulate our first main result concerning the Green's function of the resolvent.  

\begin{theorem}\label{thm.main}
Let $\gamma\in(0,+\infty)$ and $W=W^*\in L^\infty(\R_+^2,\C^{2\times 2})$. For any $z = u + iv \in \C \setminus \R$, the following hold:

\noindent{\rm (i)} The resolvent operator $(\mathscr{D}_{b,W}-z)^{-1}$ admits an integral kernel that is continuous away from the diagonal. 

\noindent{\rm (ii)} For any $c\in(0,1)$, there exists a constant $C_{\gamma,c}>0$ such that for all $x,x'\in\R_+^2$,
\[
\left|(\mathscr{D}_{b,W}-z)^{-1}(x,x')\right|\leq C_{\gamma,c}\,\frac{\langle u\rangle^6}{\min(1,|v|)}\,  \frac{e^{-c \, |v|\, |x-x'|}}{|x-x'|}\,,
\]
where $\langle u\rangle \equiv (1+|u|^2)^{1/2}$. 
\end{theorem}

\vspace{0.5cm} 
\begin{remark}
	The above estimate breaks down as $\gamma$ approaches the zigzag cases (that is, $\gamma=0$ or $\gamma=+\infty$) as is clear from Proposition~\ref{prop.free} and our proof strategy using estimates of the resolvent kernel for $b=0$ and $W=0$. 
\end{remark}
\begin{remark} \label{remark.tigre} Let $\mathrm{D}_{b,W}$ denote the ``bulk'' self-adjoint operator defined as in \eqref{tigre9} but on the whole plane $\R^2$. It is known from \cite{Cornean_2023}  that Theorem \ref{thm.main} also holds for the bulk operator with $W=0$ and $b=0$. Note that our perturbative proof strategy covers the case of a non-zero bounded $W$ and a non-zero $b$.
\end{remark}

Our second main result is related to the smooth functions of $\mathscr{D}_{b,W}$ and $\mathrm{D}_{b,W}$. 

\begin{theorem}\label{thm-tigre}
 Let $\gamma\in(0,+\infty)$ and $W=W^*\in L^\infty(\R_+^2,\C^{2\times 2})$. Let $F\in \mathscr{S}(\R)$ be any real-valued function in the Schwartz class. Then, the following properties hold:

 \noindent{\rm (i)} The operator $F(\mathscr{D}_{b,W})$ defined via functional calculus has a jointly continuous integral kernel. Moreover, for every $m\geq 1$, there exists $C_m>0$ such that for all $x = (x_1,x_2)\in\R_+^2$ and $x' = (x_1',x_2')\in\R_+^2$,
\[
\big | F(\mathscr{D}_{b,W})(x,x')\big | \leq C_m \langle x-x'\rangle^{-m}\,.
\]
The same holds true for the bulk operator $F(\mathrm{D}_{b,W})$. 

\noindent{\rm (ii)} For all $m\geq 1$, there exists $C_m>0$ such that for all $x = (x_1,x_2) \in \R^2_+$, we have 
\[
\big | F(\mathscr{D}_{b,W})(x,x)-F(\mathrm{D}_{b,W})(x,x)\big | \leq C_m \langle x_2\rangle^{-m}\,.
\]
\end{theorem}

\subsection{The bulk-edge correspondence for the non-zigzag cases} Let us give a quick description of the strategy from \cite{Cornean_2023} (which treats the infinite mass case $\gamma=1$ and it was inspired by the non-relativistic version of the problem considered in \cite{cornean2024orbital}). This approach can be repeated \textit{ad litteram} if the integral kernels of the resolvents satisfy the estimates from Theorem \ref{thm.main} and Theorem~\ref{thm-tigre}.  

Let us assume that the perturbation $W$ is $\Z^2$-periodic and that the resolvent set of the bulk operator $\mathrm{D}_{b,W}$ contains two disjoint real intervals 
\[
[e_1,e_2]\cup [e_3,e_4] \subset \R \setminus \sigma(\mathrm{D}_{b,W})\,, \quad \text{with } e_1<e_2<e_3<e_4\,.
\]
This configuration is possible, for instance, if $b\neq 0$ and the norm of $W$ is sufficiently small.

Let $F \in C_c^\infty(\R; [0,1])$ be a smooth cutoff function such that 
\begin{equation*}
F(x) = 
\begin{cases}
  1 & \text{if } e_2 \leq x \leq e_3\,, \\
  0 & \text{if } x \leq e_1 \text{ or } x \geq e_4\,.
\end{cases}
\end{equation*}
Hence $F'$ is supported in the union of the gaps of the bulk operator and $\Pi_{b,W}:=F(\mathrm{D}_{b,W})$ is a spectral projection on the interval $[e_2,e_3]$.  Since we work with the Landau gauge we define the magnetic translations as 
$$(\tau_{b,\ell} f)(x)=e^{-i\, b\,  x_1\ell_2}\, f(x-\ell),\quad \forall \ell=(\ell_1,\ell_2)\in \Z^2. $$
Then $\mathrm{D}_{b,W}$ commutes with all the magnetic translations, and so does $\Pi_{b,W}$. This implies that the diagonal value 
$$\R^2\ni x\mapsto \Pi_{b,W}(x,x)\in \C^{2\times 2}$$
is $\Z^2$ periodic. Now let us denote by $\chi_L$ the indicator function of the strip $\{0\leq x_1\leq 1,\, 0\leq x_2\leq L\}$ with $0<L\leq \infty$. Using the smooth cutoff $F$, we define a regularized version of the edge integrated density of states (IDOS) as
\[
\rho_L(b) := \frac{1}{L} \operatorname{Tr}\big(\chi_L\, F(\mathscr{D}_{b,W})\big) = \frac{1}{L}\int_0^1 \mathrm{d}x_1 \int_0^L \mathrm{d}x_2 \, \operatorname{tr}_{\C^2} F(\mathscr{D}_{b,W})(x,x)\,.
\]
Using the periodicity of $\Pi_{b,W}(x,x)$, the estimate in Theorem \ref{thm-tigre}(ii) and the identity  $\Pi_{b,W} = F(D_{b,W})$, we obtain 
\begin{equation}\label{tigre50}\lim_{L\to\infty}\rho_L(b)=\int_{[0,1]^2} {\rm tr}_{\C^2} \Pi_{b,W}(x,x)\, \mathrm{d}x=: I_\infty(b),
\end{equation}
where the quantity $I_\infty(b)$ is the integrated density of states of the bulk projection $\Pi_{b,W}$. Reasoning as in \cite[Section IV.B]{Cornean_2023} one may show the following ``commutation'' identity: 
\begin{equation}\label{tigre51}
2\pi \, \partial_b\lim_{L\to\infty}\rho_L(b)
=2\pi \, \lim_{L\to\infty}\partial_b\rho_L(b).
\end{equation}
The left hand side of \eqref{tigre51} is just $2\pi\, I_\infty'(b)$, which equals the (integer) Chern character of the bulk projection \cite{MR4310815}. The right hand side of \eqref{tigre51} can be handled exactly like in \cite[Section IV.C]{Cornean_2023} which leads to 
\begin{equation}\label{hhh4}
2\pi \, \lim_{L\to\infty}\partial_b\rho_L(b)=-2\pi {\rm Tr}\big ( \chi_\infty\, i[\mathscr{D}_{b,W},X_1]\, F'(\mathscr{D}_{b,W})\big )=-2\pi\,   {\rm Tr}\big ( \chi_\infty\,  \sigma_1\, F'(\mathscr{D}_{b,W})\big ),
\end{equation}
which is the edge conductance { and it can be shown as in \cite[Appendix B]{cornean2024orbital} to be also integer-valued, determined by the behavior of the edge states of $\mathscr{D}_{b,W}$ in the gaps where $F'$ is supported}. The equality between these two integer-valued quantities is what is meant here by bulk-edge correspondence. We note that when $W=0$, one can directly prove this equality by a careful investigation of the behavior of the edge states of the purely magnetic edge Dirac--Landau operator \cite{MR4769230}.  

\begin{remark}[What can go wrong in the zigzag case?] Let us assume that $\gamma\in \{0,\infty\}$ and also $b=0=W$. Then $\mathscr{D}_{0,0}$ has a zero-mode, \emph{i.e.} an eigenvalue of infinite multiplicity sitting at zero. Its corresponding spectral projection $\Pi_0$ has an integral kernel which up to some constants is given by 
$$\frac{1}{\big (i(x_1-x_1')+(x_2+x_2')\big )^2}.$$
This singularity near the edge remains true when $b\neq 0$, which shows that the proof method used in \cite{Cornean_2023}, based on local integrated density of states, fails to work.  
\end{remark}
\subsection{Structure of the paper}

In Section~\ref{sec:resolvent-free}, we compute the resolvent kernel of the unperturbed Dirac operator in the case $W=0$ and $b=0$, serving as a model to understand the behavior near the boundary. Section~\ref{resolventW} extends the analysis to the case $W\neq 0$ but $b=0$, where we establish continuity and localization properties of the resolvent kernel for all non-zigzag boundary conditions. Section~\ref{resolventmagnetic} introduces the magnetic field  through gauge invariant magnetic perturbation theory combined with prior estimates to control its contribution. These results deliver the proof of Theorem~\ref{thm.main}. Finally, Section~\ref{sec:functions} is devoted to proving Theorem~\ref{thm-tigre}, where we derive continuity and localization estimates for smooth functions of the edge and bulk operators using the Helffer--Sj\"ostrand formula. 

In the appendix, we collect three technical results used throughout the paper. Lemma~\ref{lem.Simon} provides sufficient conditions for constructing continuous integral kernels for operators of the form $A_1^* B A_2$ where the $A_j$'s have nice integral kernels and $B$ is a bounded operator. This is  a key tool for handling the perturbative terms in the resolvent expansions. Lemma~\ref{lem.CT} states a Combes--Thomas-type estimate. Finally, Lemma~\ref{lemma-pim} establishes exponential decay and continuity properties for composition of integral kernels.

\section{The resolvent kernel for \texorpdfstring{$W=0$}{W=0} and \texorpdfstring{$b=0$}{b=0}}
\label{sec:resolvent-free}

Recall that we assume $0<\gamma<+\infty$.
If $W=0$ the operator is invariant under translations in the $x_1$-direction, hence using the partial Fourier transform in the $x_1$-direction we have 
\[
	\mathscr{D}_{b,0}\equiv\mathscr{D}_{b} = \int_\R^\oplus\mathscr{D}_{b}(\xi)\, {\rm d}\xi\,,
\]
where the $\mathscr{D}_{b}(\xi)$'s are the one-dimensional magnetic Dirac fiber operators of $\mathscr{D}_{b}$ acting as
	\[\mathscr{D }_{b}(\xi)=\begin{pmatrix}
	0& \xi-\partial_2+bx_2\\
	\xi+\partial_2+bx_2&0
\end{pmatrix}\,,\] 
	 on 
	\[\mathrm{Dom}(\mathscr{D}_{b}(\xi))=\{\psi\in H^1({\R}_+,\C^2) : x_2\psi\in L^2(\mathbb{R}_+)\,, \psi_2(0)=\gamma\psi_1(0)\}\,,\]
    see, \emph{e.g.}, \cite{barbaroux:hal-02889558, MR4769230}. 
In order to obtain the desired bounds for the resolvent we will start by computing the integral kernel for the case $W=b=0$. 

\subsection{The resolvent kernel of the fiber operator}
Let us first investigate the corresponding resolvent kernel of the fiber operator on the whole line. This is the content of the following elementary statement. A detailed proof, based on the Fourier transform, is provided in Appendix \ref{app:proof_KR}.

\begin{lemma}\label{lem.lemma2_1}
Let $\mathscr{D}^\R_{0}(\xi)$ denote the Dirac operator defined on the whole real line acting as $\mathscr{D}_{0}(\xi)$. 
	Then, the kernel of the resolvent $(\mathscr{D}^\R_{0}(\xi)-i)^{-1}$ is $(x_2,x_2')\mapsto K^\R_{0,\xi}(x_2-x_2')$ where
	\begin{equation}\label{eq.KR}
    \begin{aligned}
	K^\R_{0,\xi}(x_2)&=\mathscr{F}^{-1}_{k \to x_2}\left(\frac{i}{k^2+\xi^2+1}\begin{pmatrix}
		1&-k-i\xi\\
		k-i\xi&1
	\end{pmatrix}\right)(x_2)\,\\
    &=\frac{1}{2\langle \xi\rangle }e^{-\langle \xi\rangle |x_2|}\begin{pmatrix}
		i&\xi+\langle \xi\rangle \,\sign(x_2 )\\
		\xi-\langle \xi\rangle \, \sign(x_2)&i
	\end{pmatrix}\,.
    \end{aligned}
    \end{equation}
    Here, $\mathscr{F}_{x_2 \to k}f(k) = \int_{\R} f(x_2) e^{-i x_2 k}\mathrm{d}x_2$ is the Fourier transform, where $k$ is the Fourier variable associated with the spatial coordinate $x_2$.
\end{lemma}
Now we present our results for the resolvent kernel of the free Dirac operator $\mathscr{D}_{0}(\xi)-i$ on the half-line.

\begin{proposition}
Let $\gamma\in[0,+\infty]$ and  
\begin{equation}\label{eq.alpha}
	\alpha(\xi)=\frac{\gamma+i(\langle \xi\rangle +\xi)}{1+i\gamma(\langle \xi\rangle +\xi)}\,.
		\end{equation}
        We define 
	\begin{equation}\label{eq.K0xilambda}K_{0,\xi}(x_2,x_2')=K^\R_{0,\xi}(x_2-x_2')+\alpha(\xi){K}^\R_{0,\xi}(x_2+x_2')\sigma_1\,,\quad x_2,x_2'\geq 0\,.\end{equation}
    For each $\Psi\in C^\infty_0(\R_+,\C^2)$, we consider

    \[\Phi(x_2)=\int_{\R_+} K_{0,\xi}(x_2,x_2')\Psi(x_2')\mathrm{d}x_2'\,,\quad x_2\geq 0\, .\]
	Then, we have $\Phi=[\Phi_1,\Phi_2]^\top\in\mathscr{S}(\overline{\R}_+,\C^2)$, 
	\begin{equation}\label{stock1} (\mathscr{D}_{0}(\xi)-i)\Phi=\Psi\,, \quad \text{and}\quad\Phi_2(0)=\gamma \, \Phi_1(0).
    \end{equation}
    In particular, $K_{0,\xi}(x_2,x_2')$ is the integral kernel of $(\mathscr{D}_{0}(\xi)-i)^{-1}$. 
\end{proposition}

\begin{proof}
	Let $\Psi\in C^\infty_0(\R_+,\C^2)$. Let us consider 
	\[\Phi_0(x_2)=K^\R_{0,\xi}\ast   \Psi(x_2)=\int_{\R_+} K^\R_{0,\xi}(x_2-x_2')\Psi(x_2')\mathrm{d}x_2'\,.\]
	By Lemma \ref{lem.lemma2_1}, this belongs to the Schwartz class since $K^\R_{0,\xi}\in L^\infty(\R)\cap L^1(\R,e^{|x_2|/2}\mathrm{d} x_2)$ and $\Psi$ has compact support. We have
	\[(\mathscr{D}^\R_{0}(\xi)-i)\Phi_0=\Psi\,.	\]
	Let us construct a function $\Phi$ satisfying \eqref{stock1}. We must have, as a differential equation on $\R_+$,
	\[(\mathscr{D}_{0}(\xi)-i)(\Phi-\Phi_0)=0\,.\]
	The linear and homogeneous first-order differential equation $(\mathscr{D}_{0}(\xi)-i)u=0$ has a general solution $u=[u_1,u_2]^\top $ expressed as a linear combination with two constants $\alpha,\beta \in\C$ such that
	\[u_1(x_2)=i(\langle \xi\rangle -\xi) \, \alpha\,  e^{\langle \xi\rangle  x_2}-i(\langle \xi\rangle +\xi)\,  \beta \, e^{-\langle \xi\rangle  x_2}\,, \qquad u_2(x_2)=\alpha\,  e^{\langle \xi\rangle  x_2}+\beta \, e^{-\langle \xi\rangle  x_2}\,.\]
    Thus $\Phi-\Phi_0$ must be of this form. 
	To ensure that $\Phi$ does not grow at infinity, we must have $\alpha=0$. This means that 
	\[\Phi(x_2)=\Phi_0(x_2)+\beta e^{-\langle \xi\rangle  x_2}v_\xi\,,\quad v_\xi= \begin{pmatrix}
		-i(\langle \xi\rangle +\xi)\\
		1
	\end{pmatrix}\,.\]
	For $\gamma \in [0, +\infty)$, we then determine $\beta$ such that the boundary condition $\Phi_2(0)=\gamma \Phi_1(0)$ from \eqref{stock1} is satisfied. At $x_2=0$, we have $\Phi(0)=\Phi_0(0)+\beta v_\xi$.
	Introducing 
		\begin{equation}\label{eq.beta}
		\beta_1:=\frac{\gamma}{1+i(\langle \xi\rangle +\xi)\gamma}\,,\quad\text{and}\quad \beta_2:=-\frac{1}{1+i(\langle \xi\rangle +\xi)\gamma}\,,
	\end{equation}
    we find that
		\[\beta=\frac{\Phi_{0,1}(0)\gamma-\Phi_{0,2}(0)}{1+i(\langle \xi\rangle +\xi)\gamma}=\beta_1\Phi_{0,1}(0)+\beta_2\Phi_{0,2}(0)\,,\]
    where $\phi_0=[\phi_{0,1},\phi_{0,2}]^\top $.
	Denoting by $[a\,|\, b]$ the matrix formed by the column vectors $a$ and $b$, we have, by Lemma \ref{lem.lemma2_1},
	\[\Phi(x_2)=\Phi_0(x_2)+e^{-\langle \xi\rangle  x_2}[\beta_1 v_\xi\, |\, \beta_2v_\xi] \Phi_0(0)\,.\]
    This ensures that $\Phi$ satisfies the boundary condition from \eqref{stock1}.
	Let us analyze the contribution of the second term on the right-hand side. Since $x_2' \geq 0$, we have $|-x_2'| = x_2'$ and $\sign(-x_2') = -1$, yielding:
	\[\begin{split}
		e^{-\langle \xi\rangle  x_2}[\beta_1 v_\xi\, |\, \beta_2v_\xi] \Phi_0(0)
		&=e^{-\langle \xi\rangle  x_2}\int_{\R_+} [\beta_1 v_\xi\, |\, \beta_2v_\xi] K^\R_{0,\xi}(-x_2')\Psi(x_2')\mathrm{d}x_2'\\
		&=\frac{1}{2\langle \xi\rangle }\int_{\R_+} e^{-\langle \xi\rangle (x_2+x_2')}[\beta_1 v_\xi\, |\, \beta_2v_\xi] \begin{pmatrix}
			i&\xi-\langle \xi\rangle \\
			\xi+\langle \xi\rangle &i
		\end{pmatrix}\Psi(x_2')\mathrm{d}x_2'\,.
	\end{split}\]
    A straightforward computation using the identity $\langle \xi\rangle ^2=\xi^2+1$ shows that
	\begin{equation*}
\begin{split}			 [\beta_1 v_\xi\, |\, \beta_2v_\xi] \begin{pmatrix}
				i &\xi-\langle \xi\rangle \\
				\xi+\langle \xi\rangle &i   
			\end{pmatrix}
			&=(1+i(\langle \xi\rangle +\xi)\gamma)^{-1}[\gamma v_\xi\, |\, -v_\xi]\begin{pmatrix}
				i &\xi-\langle \xi\rangle \\
				\xi+\langle \xi\rangle &i   
			\end{pmatrix}\\
			&=\alpha(\xi)\begin{pmatrix}
				i &\xi+\langle \xi\rangle \\
				\xi-\langle \xi\rangle &i   
			\end{pmatrix}\sigma_1\,,
	\end{split}
	\end{equation*}
	where $\alpha(\xi)$ is given by \eqref{eq.alpha}. In view of \eqref{eq.KR}, this gives the expected result for the integral kernel when $\gamma \in [0, +\infty)$.
	
	To conclude, let us address the case $\gamma = +\infty$, which corresponds to the Dirichlet boundary condition on the first component ($\psi_1(0) = 0$). While the algebraic derivation above assumes $\gamma < +\infty$ to well define the fractions $\beta_1$ and $\beta_2$, the integral kernel for $\gamma = +\infty$ can be rigorously recovered by taking the limit $\gamma \to +\infty$ in the definition of $\alpha(\xi)$, which yields $\alpha(\xi) = -i(\langle \xi \rangle - \xi)$.
\end{proof}

        For future reference, we establish the following technical result.

\begin{lemma}\label{lem.alpha}
	Let $\gamma \in (0, +\infty)$. Recall that $\alpha(\xi)$ is given by \eqref{eq.alpha}. We have, for all $\xi\in\R$,
	\[
\min\left(\gamma,\gamma^{-1}\right)\leq|\alpha(\xi)|\leq\max\left(\gamma,\gamma^{-1}\right)\quad\text{and}\quad
	\alpha(-\xi)=(\overline{\alpha(\xi)})^{-1}\,.
	\]
	Moreover, there exists $C(\gamma)>0$ such that, for all $\xi\in\R$,
	\begin{equation}\label{eq.alpha'}
	|\alpha'(\xi)|\leq C(\gamma)\langle \xi\rangle^{-2}\,.	
	\end{equation}
\end{lemma}
\begin{proof}
	First, we prove the bounds on $|\alpha(\xi)|$. Let us set $x(\xi) = \langle \xi \rangle + \xi$. Since $\langle \xi \rangle = \sqrt{\xi^2+1} > |\xi|$, we have $x(\xi) > 0$ for all $\xi \in \R$. We can write $\alpha(\xi) = f(x(\xi))$ where $f(x) = \frac{\gamma + i x}{1 + i \gamma x}$. 
	The squared modulus of $f(x)$ is given by
	\[
	|f(x)|^2 = \frac{\gamma^2 + x^2}{1 + \gamma^2 x^2}\,.
	\]
	Consider the function $h(s) = \frac{\gamma^2 + s}{1 + \gamma^2 s}$ for $s = x^2 \geq 0$. Its derivative is 
	\[
	h'(s) = \frac{1 - \gamma^4}{(1 + \gamma^2 s)^2}\,.
	\]
	Since the sign of $h'(s)$ is constant and depends only on whether $\gamma < 1$ or $\gamma > 1$, the function $h(s)$ is monotonic on $[0, +\infty)$. Thus, its values are strictly bounded between $h(0) = \gamma^2$ and its limit at infinity $\lim_{s \to +\infty} h(s) = \gamma^{-2}$. Taking the square root, we deduce that $|\alpha(\xi)| = \sqrt{h(x(\xi)^2)}$ is strictly bounded between $\min(\gamma, \gamma^{-1})$ and $\max(\gamma, \gamma^{-1})$, establishing the first part of the statement.
	
	Next, we detail the proof of the derivative bound \eqref{eq.alpha'}. Notice that we can write
	\[\alpha(\xi)=g(x(\xi))\,,\quad \text{where} \quad g(x)=\frac{1}{\gamma}+\frac{\gamma-\gamma^{-1}}{1+i\gamma x}\,.\]
	To establish the bound on $g'$, we compute its derivative:
	\[ 
	g'(x) 
	= (\gamma - \gamma^{-1}) \frac{-i\gamma}{(1 + i\gamma x)^2} = \frac{i(1 - \gamma^2)}{(1 + i\gamma x)^2} \,. 
	\]
	Taking the modulus yields:
	\[ 
	|g'(x)| = \frac{|1 - \gamma^2|}{|1 + i\gamma x|^2} = \frac{|1 - \gamma^2|}{1 + \gamma^2 x^2} \,. 
	\]
	Using the elementary inequality $1 + \gamma^2 x^2 \geq \min(1, \gamma^2)(1 + x^2) = \min(1, \gamma^2)\langle x \rangle^2$, we can bound $|g'(x)|$ as follows:
	\[ 
	|g'(x)| \leq \frac{|1 - \gamma^2|}{\min(1, \gamma^2)} \frac{1}{1 + x^2} = \tilde{C}(\gamma) \langle x\rangle^{-2} \,, 
	\]
	where $\tilde{C}(\gamma) = |1 - \gamma^2| \max(1, \gamma^{-2}) = \max(\gamma^2,\gamma^{-2})-1$. 
	
	Furthermore, the derivative of the inner function is $x'(\xi) = \frac{\xi}{\langle \xi \rangle} + 1 = \frac{x(\xi)}{\langle \xi \rangle}$. 
	Thus, using the chain rule, we obtain for $\xi \geq 0$ (where $x(\xi) \geq \langle \xi \rangle \geq 1$):
	\[
	|\alpha'(\xi)| = |g'(x(\xi))|\, x'(\xi) \leq \tilde{C}(\gamma) \frac{1}{\langle x(\xi) \rangle^2} \frac{x(\xi)}{\langle \xi\rangle} \leq \tilde{C}(\gamma) \frac{x(\xi)}{x(\xi)^2 \langle \xi \rangle} = \frac{\tilde{C}(\gamma)}{x(\xi) \langle \xi\rangle}\,.
	\]
	Since $x(\xi) \geq \langle \xi \rangle$ for $\xi \geq 0$, we conclude that $|\alpha'(\xi)| \leq \tilde{C}(\gamma)\langle \xi \rangle^{-2}$, which yields \eqref{eq.alpha'} for $\xi \geq 0$.
	
	For $\xi < 0$, we rely on the symmetry property $\alpha(\xi)=(\overline{\alpha(-\xi)})^{-1}$. Differentiating this relation with respect to $\xi$ gives:
	\[|\alpha'(\xi)|=\frac{|\alpha'(-\xi)|}{|\alpha(-\xi)|^2}\leq\max(\gamma^2,\gamma^{-2})|\alpha'(-\xi)|\,.\]
	Since $-\xi > 0$, we can apply the bound previously established to $\alpha'(-\xi)$, concluding the proof for all $\xi \in \R$.
\end{proof}

\subsection{The resolvent kernel on the upper plane}

By using the inverse Fourier transform, we notice that the integral kernel of $(\mathscr{D}_0-i)^{-1}$ is
\begin{equation}\label{eq.K0lambda}
	K_{0}(x,x'):=\frac{1}{2\pi}\int_{\R}e^{i\xi(x_1-x'_1)}K_{0,\xi}(x_2,x'_2)\mathrm{d}\xi\,,
\end{equation}
where $K_{0,\xi}$ is explicitly given by \eqref{eq.K0xilambda}. 
Observe that, when $\gamma=1$ we have that $\alpha(\xi)=1$ and we recover the expression of the kernel found in \cite{Cornean_2023} for the infinite mass case. The main result  of this subsection is the following proposition.

\begin{proposition}\label{prop.free}
	For all $c\in(0,1)$ and all $\gamma>0$, there exists  a constant $C(\gamma,c)>0$ such that for all $x,x'\in \R^2_+$ and $\lambda\in \R\setminus\{0\}$, we have
	\begin{equation}\label{pim1}|(\mathscr{D}_0-i\lambda)^{-1}(x,x')|\leq C(\gamma,c)\, \frac{e^{-c\,|\lambda|\, |x-x'|}}{|x-x'|}\,.
    \end{equation}
    Moreover, the constant $C(\gamma,c)$ diverges when either $\gamma$ goes to $0$ or $+\infty$, or when $c$ goes to $1$.
\end{proposition}

	\begin{proof}
	By noticing that (see Notation \ref{stock3}):
    $$(\mathscr{D}_0+i\lambda)^{-1}(x,x')=\Big ((\mathscr{D}_0-i\lambda)^{-1}\Big )^*(x,x')=\Big ((\mathscr{D}_0-i\lambda)^{-1}(x',x)\Big )^\dagger\, ,$$
     we can reduce the problem to $\lambda >0$. Moreover, by scaling, it is enough to consider the case $\lambda=1$; indeed, if we introduce the unitary operator $(U_\lambda f)(x)=\lambda
     \, f(\lambda \, x)$ we have 
     $$ U_\lambda^* (\mathscr{D}_0-i\lambda )U_\lambda =\lambda( \mathscr{D}_0-i)\quad \text{and}\quad(\mathscr{D}_0-i\lambda )^{-1}=\lambda^{-1}\,  U_\lambda(\mathscr{D}_0-i )^{-1} U_\lambda^*\, .$$
     At the level of integral kernels this gives 
     \begin{align*}(\mathscr{D}_0-i\, \lambda )^{-1}\psi(x)&=\int_{\R^2_+}(\mathscr{D}_0-i\, \lambda )^{-1}(x,x')\psi(x')\mathrm{d}x'
     =\lambda^{-1}\int_{\R^2_+}K_{0 }(\lambda x,y)\psi(\lambda^{-1} y)\mathrm{d}y\\
     &=\lambda\int_{\R^2_+}K_{0 }(\lambda x,\lambda x')\psi(x')\mathrm{d}x'\, .
     \end{align*}
Thus, 
\begin{equation}\label{stock5}
    (\mathscr{D}_0-i\, \lambda )^{-1}(x,x')=\lambda\, K_{0 }(\lambda x,\lambda x')\, ,
\end{equation}
which implies that it is enough to prove \eqref{pim1} for $\lambda=1$. 

From \eqref{eq.K0lambda} and \eqref{eq.K0xilambda} we have
$$K_0(x,x')=\frac{1}{2\pi}\int_\R\, e^{i\xi(x_1-x_1')}\,  K^\R_{0,\xi}(x_2-x_2') + \frac{1}{2\pi} \, I(x,x')\sigma_1$$
where 
\[I(x,x'):=\int_{\R}e^{i\xi(x_1-x'_1)}\alpha(\xi){K}^\R_{0,\xi}(x_2+x'_2)\mathrm{d}\xi\,.\]
The first term gives the integral kernel of the resolvent of the ``bulk'', which is well-known and satisfies \eqref{pim1} (see, \emph{e.g.}, \cite{Cornean_2023}), hence it suffices to show that $I(x,x')$---the ``edge'' contribution---also satisfies the required estimate.

To do so, we write $I(x,x')=\frac12F(S,T)$ with $S=x_1-x'_1$ and $T=|x_2+x'_2|=x_2+x'_2>0$. We have
\[F(S,T)=\int_{\R}e^{i\xi S-\langle\xi\rangle T}A(\xi)\mathrm{d}\xi\,,\quad A(\xi)=\frac{\gamma+i(\langle\xi\rangle+\xi)}{1+i\gamma(\langle\xi\rangle+\xi)}\begin{pmatrix}
	i\langle\xi\rangle^{-1}&\xi\langle\xi\rangle^{-1}+1\\
	\xi\langle\xi\rangle^{-1}-1&i\langle\xi\rangle^{-1}
\end{pmatrix}\,.\]
Noting that $\xi\mapsto A(\xi)$ is uniformly bounded (see Lemma \ref{lem.alpha}), we have that the function in the integral is in $L^1(\R)$ since $T>0$. 

Consider the change of variable $\xi=\sinh x$. We get
\[F(S,T)=\int_{\R}e^{iS\sinh x- T\cosh x}A(\sinh(x))\cosh(x)\mathrm{d}x\,.\]
We notice that
\begin{equation}\label{eq.A_1sinh}
A(\sinh(x))\cosh(x)=\frac{\gamma+ie^x}{1+i\gamma e^{x}}\begin{pmatrix}
	i&e^x\\
	-e^{-x}&i
\end{pmatrix}\quad\text{and}\quad\frac{\gamma+ie^x}{1+i\gamma e^{x}}=\frac{1}{\gamma}+\frac{\gamma-\gamma^{-1}}{1+i\gamma e^{x}}\,.
\end{equation}
Let us consider $\theta\in\left(-\frac\pi2,\frac\pi2\right)$ and $r>0$ such that $S=r\sin\theta$ and $T=r\cos\theta$. We notice that
\[\sinh (i\theta)=i\sin\theta\,,\quad \cosh (i\theta)=\cos \theta\,,\]
and
\[\cosh(x-i\theta)=\cosh(x)\cosh (i\theta)-\sinh(x)\sinh(i\theta)\,.\]
Thus,
\[F(S,T)=\int_{\R}e^{-r\cosh(x-i\theta)}A(\sinh(x))\cosh(x)\mathrm{d}x\,.\]
Take $\epsilon\in(0,\frac\pi2)$ and let $\theta_\epsilon=\theta-\epsilon\in(-\pi,\pi/2)$. Consider the complex translation $x\mapsto x+i\theta_\epsilon$. We can apply the Cauchy theorem to get
\[F(S,T)=\int_{\R}e^{-r\cosh(x-i\epsilon)}A(\sinh(x+i\theta_\epsilon))\cosh(x+i\theta_\epsilon)\mathrm{d}x\,,\]
so that
\[\begin{split}
|F(S,T)|&\leq\int_{\R}e^{-r\cos\epsilon\cosh(x)}|A(\sinh(x+i\theta_\epsilon))\cosh(x+i\theta_\epsilon)|\mathrm{d}x
\\&\leq e^{- r_\epsilon}\int_{\R}e^{-2r_\epsilon\sinh^2(x/2)}|A(\sinh(x+i\theta_\epsilon))\cosh(x+i\theta_\epsilon)|\mathrm{d}x\,,
\end{split}\]
where $r_\epsilon=r\cos\epsilon$.
Considering the change of variable $y=\sinh(x/2)$, or equivalently $x=\varphi(y)=2\ln(y+\sqrt{y^2+1})$, we get
\[|F(S,T)|\leq e^{- r_\epsilon}\int_{\R}e^{-2r_\epsilon y^2}|A(\sinh(\varphi(y)+i\theta_\epsilon))\cosh(		\varphi(y)+i\theta_\epsilon)|\varphi'(y)\mathrm{d}y\,.\]
Thanks to \eqref{eq.A_1sinh}, we deduce that
\[
    |\cosh(\varphi(y)+i\theta_\epsilon)A(\sinh(\varphi(y)+i\theta_\epsilon))|= \left|\frac{\gamma+ie^{\varphi(y)+i\theta_\epsilon}}{1+i\gamma e^{{\varphi(y)+i\theta_\epsilon}}}\right|\left|\begin{pmatrix}
 	i&e^{i\theta_\epsilon}e^{\varphi(y)}\\
 	-e^{-i\theta_\epsilon}e^{-\varphi(y)}&i
 \end{pmatrix}\right|.
\]
Direct computation of the spectral norm yields
\[
    \left| \begin{pmatrix}
 	i & e^{i\theta_\epsilon}e^{\varphi(y)} \\
 	-e^{-i\theta_\epsilon}e^{-\varphi(y)} & i
 \end{pmatrix} \right| = 2\cosh(\varphi(y))\leq 4\langle y\rangle^2 \,.
\]
Furthermore, setting $X = e^{\varphi(y)} \in (0, +\infty)$ and $Y = \sin\theta_\epsilon \in [-1, 1)$, we observe the algebraic identity
\[
    \left|\frac{\gamma+ie^{\varphi(y)+i\theta_\epsilon}}{1+i\gamma e^{\varphi(y)+i\theta_\epsilon}}\right|^2 
    = \frac{X^2 - 2\gamma Y X + \gamma^2}{\gamma^2 X^2 - 2\gamma Y X + 1}
    \,.
\]
Note that the numerator and the denominator are positive since $Y<1$. To obtain a uniform upper bound over $X \in (0, +\infty)$, we would like to find a constant $Z>0$ such that for all $X>0$ we have  
\[
    \frac{X^2 - 2\gamma Y X + \gamma^2}{\gamma^2 X^2 - 2\gamma Y X + 1} \leq Z.
\]
This holds if and only if the polynomial 
\[
    P(X) = Z(\gamma^2 X^2 - 2\gamma Y X + 1) - (X^2 - 2\gamma Y X + \gamma^2) = (\gamma^2 Z - 1)X^2 - 2\gamma Y(Z - 1)X + (Z - \gamma^2)
\]
satisfies $P(X) \geq 0$ on $(0, +\infty)$. For $Y \leq 0$, this non-negativity condition is trivially satisfied by choosing $Z = \max(\gamma^2, \gamma^{-2})$.

For $Y \in (0, 1)$, we analyze the reduced discriminant of $P(X)$ with respect to $X$, which must be non-positive. This requirement yields the following symmetric polynomial inequality in $Z$:
\[
    \gamma^2(Y^2 - 1)Z^2 + (\gamma^4 - 2\gamma^2 Y^2 + 1)Z + \gamma^2(Y^2 - 1) \leq 0\,.
\]
This polynomial is of the form $p(z)=a z^2 +b z +a$ with $a<0$ and $b>0$. Its discriminant is $d=b^2-4a^2$. 
We have 
\[
\frac{b}{-a}=\frac{\gamma^4 - 2\gamma^2 Y^2 + 1}{\gamma^2(1 - Y^2)} = 2 + \frac{(\gamma - \gamma^{-1})^2}{1 - Y^2}\geq 2
\]
hence $b^2/a^2\geq 4$ and $d\geq 0$.  
Then $p(z)$ is positive only between its two (positive) roots. The largest root is 
\[
(b+\sqrt{d})/(-2a)\leq -b/a= 2 + \frac{(\gamma - \gamma^{-1})^2}{1 - Y^2}\leq \frac{4\max(\gamma^2, \gamma^{-2})}{1 - Y^2}\,.
\]
Any $Z$ larger than the right hand side obeys $p(Z)\leq 0$ and thus $P(X)\geq 0$. Thus if $Y\in [-1,1)$ we may combine the bounds and choose $Z=\frac{4\max(\gamma^2, \gamma^{-2})}{1 - \chi_{(0,1)}(Y)Y^2}$. This upper bound can explode only when $Y$ is close to $1$. 
Recalling that $\theta_\epsilon = \theta - \epsilon \leq \pi/2 - \epsilon$, the condition $Y > 0$ implies $\theta_\epsilon>0$. Thus, $\cos(\theta_\epsilon) \geq \sin(\epsilon)$, which in turn gives $1 - Y^2 = \cos^2(\theta_\epsilon) \geq \sin^2(\epsilon)$. We conclude that the modulus is bounded uniformly in $\varphi(y)$ by
\[
    \left|\frac{\gamma+ie^{\varphi(y)+i\theta_\epsilon}}{1+i\gamma e^{\varphi(y)+i\theta_\epsilon}}\right| \leq \frac{2\max(\gamma, \gamma^{-1})}{\sin(\epsilon)}
\]
and
\[ |\cosh(\varphi(y)+i\theta)A(\sinh(\varphi(y)+i\theta))|\leq 8\frac{\max(\gamma,\gamma^{-1})}{\sin(\epsilon)}\langle y\rangle^2=\tilde C(\gamma,\epsilon)\langle y\rangle^2\,.\]
We also recall that $\varphi'(y)=\frac{2}{\sqrt{y^2+1}}$. Hence, for all $(S,T)\in\R\times\R_+$,
\[|F(S,T)|\leq 2\tilde C(\gamma,\epsilon) e^{- r_\epsilon}\int_{\R}e^{-2r_\epsilon y^2}\langle y\rangle\mathrm{d}y\leq \hat C(\gamma,\epsilon) e^{-r_\epsilon}\left( \frac{1}{\sqrt{r_\epsilon}}+\frac{1}{r_\epsilon}\right)\,.\]
Therefore, for all $c\in(0,1)$, there exists $C(\gamma,c)>0$ such that, for all $(S,T)\in\R\times\R_+$, we have
\[|F(S,T)|\leq \frac{C(\gamma,c)}{r}e^{-c\,  r}\,,\mbox{ where } r=\sqrt{S^2+T^2}\geq |x-x'|\,,\]
and this shows that $I(x,x')$ satisfies the estimate in \eqref{pim1}. 
\end{proof}

\section{The resolvent kernel for \texorpdfstring{$W\neq 0$}{W != 0} and \texorpdfstring{$b=0$}{b=0}}
\label{resolventW}
The aim of this section is to prove the following proposition.

\begin{proposition}\label{prop.resolventWz}
	Let $W\in L^\infty(\R_+^2,\C^{2\times 2})$ be a two by two Hermitian matrix-valued function. Then, 
	the operator $(\mathscr{D}_0+W-i\lambda )^{-1}$ with $\lambda>0$ has an integral kernel that is continuous away from the diagonal. Moreover, for all $c\in(0,1)$ and all $\gamma\in(0,+\infty)$, there exists $C_0>0$ such that, for all $(x,x')\in\R_+^2\times\R^2_+$, and all $\lambda>0$,
	\begin{equation}\label{pim9}|(\mathscr{D}_0+W-i\lambda)^{-1}(x,x')|\leq C_0\max(1,\lambda^{-1})|x-x'|^{-1}e^{-c\lambda|x-x'|}\,.\end{equation}
	Moreover:
    
      \noindent{\rm (i)}  For $j\geq 1$, we have that $ (\mathscr{D}_0+W-i\lambda )^{-j}(x,x')$ is continuous outside the diagonal  and is exponentially localized around it;
      
      \noindent{\rm (ii)} $ (\mathscr{D}_0+W-i\lambda )^{-2}(x,x')$ has a logarithmic singularity on the diagonal;
      
      \noindent{\rm (iii)} For $j\geq 3$, $ (\mathscr{D}_0+W-i\lambda )^{-j}(x,x')$ is jointly continuous. 
    
    \end{proposition}

    \vspace{0.5cm}

Before giving the proof, let us outline the main strategy behind it. The starting point is the following formula which follows from six iterations of the second resolvent identity: 
\begin{align}\label{pim2}
(\mathscr{D}_0+W-i\lambda)^{-1}
		=&\sum_{j=0}^5 (-1)^j[(\mathscr{D}_0-i\lambda)^{-1}W]^j(\mathscr{D}_0-i\lambda)^{-1}\\
        &\quad +[(\mathscr{D}_0-i\lambda)^{-1}W]^3(\mathscr{D}_0+W-i\lambda)^{-1}[W(\mathscr{D}_0-i\lambda)^{-1}]^3\,.\nonumber 
	\end{align}
The right hand side consists of seven terms. The first six (included in the sum) have an obvious integral kernel and we will show that as $j$ grows, the regularity of the kernels of these terms improve. In particular, if $j\geq 3$, the kernels will be continuous even on the diagonal. The integral kernel of the last term in \eqref{pim2} is not explicit. However, we will construct it rigorously in Lemma \ref{lem.Simon}, relying only on the boundedness of the perturbed resolvent and the continuity of the integral kernels of $[(\mathscr{D}_0-i\lambda)^{-1}W]^3$ and $[W(\mathscr{D}_0-i\lambda)^{-1}]^3$---this explains the necessity of iterating six times in \eqref{pim2}. The exponential localization of this term is proved with the method used in Lemma \ref{lemma-pim} and Remark \ref{remark-pim}. 

Next we argue that it is enough to prove the proposition for $\lambda\ge 1$. Indeed, when $\lambda\in(0,1)$, we use the first resolvent formula with $z=i\lambda$ to obtain, writing $\mathscr{D}\equiv \mathscr{D}_0+W$, that
\begin{equation}\label{pim8}
(\mathscr{D}-z)^{-1}=\sum_{j=0}^5(z-i)^j(\mathscr{D}-i)^{-j-1}+(z-i)^{6}(\mathscr{D}-i)^{-3}(\mathscr{D}-z)^{-1}(\mathscr{D}-i)^{-3}\,.
\end{equation} 
The term $\sum_{j=0}^5(z-i)^j(\mathscr{D}-i)^{-j-1}$ is covered by the case with $\lambda=1$. The last term on the right hand side of  \eqref{pim8} can be treated with Lemma \ref{lemma-pim} which leads to
\[|(\mathscr{D}-i)^{-3}(\mathscr{D}-z)^{-1}(\mathscr{D}-i)^{-3}(x,x')|\leq C\lambda^{-1}e^{-\lambda c|x-x'|}\,.\]	
This proves \eqref{pim9} for $0<\lambda<1$. Next, we discuss the last three points of the proposition when $0<\lambda<1$. By squaring \eqref{pim8}, the most singular term is $(\mathscr{D}-i)^{-2}$ which has a logarithmic singularity at the diagonal. Raising \eqref{pim8} to power $3$ or higher will generate  continuous integral kernels.

From now on we will assume that $\lambda\geq 1$. We start by stating two results concerning the sum in \eqref{pim2}, namely bounds for the integral kernels of
$(\mathscr{D}_0-i\lambda)^{-1}W(\mathscr{D}_0-i\lambda)^{-1}$ and 
 $ (\mathscr{D}_0-i\lambda)^{-1}W(\mathscr{D}_0-i\lambda)^{-1}W(\mathscr{D}_0-i\lambda)^{-1}$.
\begin{lemma}[Composing two resolvents]\label{lem.A}
The integral kernel $\mathbb{A}$ of $$A:=(\mathscr{D}_0-i\lambda)^{-1}W(\mathscr{D}_0-i\lambda)^{-1}$$ is continuous outside the diagonal $x=x'$. For all $c\in(0,1)$, there exists $C>0$ such that, for all $(x,x')\in\R^2_+$, and $\lambda\ge 1$,
\begin{equation}\label{pim11}|\mathbb{A}(x,x')|\leq C\ln(2+|x-x'|^{-1})e^{-c\lambda |x-x'|}\,.\end{equation}
	\end{lemma}

    \vspace{0.2cm}
    
\begin{lemma}[Composing three resolvents]\label{lem.A2}
The integral kernel of $AW(\mathscr{D}_0-i\lambda)^{-1}$ is continuous. Moreover, for all $c\in(0,1)$, there exists $C>0$ such that, for all $(x,x')\in\R^2_+\times\R^2_+$ and $\lambda\geq 1$,
\begin{equation}\label{pim12}|AW(\mathscr{D}_0-i\lambda)^{-1}(x,x')|\leq Ce^{-c\lambda|x-x'|}\,.\end{equation}	
	\end{lemma}

    \vspace{0.2cm}

The proofs of the above two lemmas will be given later on in the section.

\vspace{0.2cm}

    \begin{proof}[Proof of Proposition \ref{prop.resolventWz}]

From Lemmas \ref{lem.A} and \ref{lem.A2}, we see that the most singular term in the sum from \eqref{pim2} is the one with $j=0$ which was treated in the previous section. Thus we have
\[\left|\sum_{j=0}^5 (-1)^j[(\mathscr{D}_0-i\lambda)^{-1}W]^j(\mathscr{D}_0-i\lambda)^{-1}(x,x')\right|\leq C|x-x'|^{-1}e^{-c\lambda|x-x'|}\,,\]
and this integral kernel is continuous outside $x=x'$.

To establish that $(\mathscr{D}_0+W-i\lambda)^{-1}$ has an integral kernel with the announced properties, we still need to analyze the following operator
\begin{multline*}
[(\mathscr{D}_0-i\lambda)^{-1}W]^3(\mathscr{D}_0+W-i\lambda)^{-1}[W(\mathscr{D}_0-i\lambda)^{-1}]^3\\
=[AW(\mathscr{D}_0-i\lambda)^{-1}]W(\mathscr{D}_0+W-i\lambda)^{-1}W[AW(\mathscr{D}_0-i\lambda)^{-1}]\,.
\end{multline*}
We are in position to apply Lemma \ref{lem.Simon}. Indeed, consider  \[\mathbb{A}_1(x,x')=[AW(\mathscr{D}_0-i\lambda)^{-1}(x',x)]^\dagger\,,\quad \mathbb{A}_2(x,x')=AW(\mathscr{D}_0-i\lambda)^{-1}(x,x')\,,\]
and the bounded operator $B=W(\mathscr{D}_0+W-i\lambda)^{-1}W$. By using the continuity of $\mathbb{A}_j$ and the bound $|\mathbb{A}_j(x,x')|\leq Ce^{-c\lambda|x-x'|}$, we can apply the dominated convergence theorem and we have
\[\lim_{x'\to x'_0}\int_{\R^2_+}|\mathbb{A}_j(x,x')-\mathbb{A}_j(x,x'_0)|^2\mathrm{d}x=0\,.\]
Therefore the assumptions of Lemma \ref{lem.Simon} are satisfied. We deduce that the operator $[(\mathscr{D}_0-i\lambda)^{-1}W]^3(\mathscr{D}_0+W-i\lambda)^{-1}[W(\mathscr{D}_0-i\lambda)^{-1}]^3$ has a continuous integral kernel. 

In order to prove its exponential localization near the diagonal we can use Lemma \ref{lemma-pim} and Remark \ref{remark-pim} with $v=\lambda$, $b=0$ and $\mu=c\, \lambda$. 

We prove Proposition \ref{prop.resolventWz}(ii) and (iii) by applying Lemmas \ref{lem.A} and \ref{lem.A2} to the appropriate power of \eqref{pim2}.

    \end{proof}

\subsection{Proof of Lemma \ref{lem.A}}

We adapt some methods developed for the so-called kernels of potential type (or polar kernels) \cite[Section 15.4]{vladimirov1971equations}. Let us write $$K_{0,\lambda}\equiv (\mathscr{D}_0-i\lambda)^{-1}(x,x').$$
    We have
	\[\mathbb{A}(x,x')=\int_{\R^2_+}K_{0, \lambda}(x,y)W(y)K_{0,\lambda}(y,x')\mathrm{d}y\,.\]
	\noindent{\bf Proof of the bound in \eqref{pim11}.} From Proposition \ref{prop.free}, we get
		\[\begin{split}|\mathbb{A}(x,x')|&\leq\|W\|_\infty\int_{\R^2_+}|K_{0, \lambda}(x,y)||K_{0,\lambda}(y,x')|\mathrm{d}y\\
		&\leq C\|W\|_\infty\int_{\R_+^2}|x-y|^{-1}|x'-y|^{-1}e^{-c\lambda(|x-y|+|x'-y|)}\mathrm{d}y\\
		&\leq C\|W\|_\infty e^{-(c-\epsilon)\lambda|x-x'|}\int_{\R_+^2}|x-y|^{-1}|x'-y|^{-1} e^{-\epsilon\lambda(|x-y|+|x'-y|)}\mathrm{d}y\,,
		\end{split}\]
		for all $\epsilon\in(0,c)$ and where the triangle inequality was used to get the last estimate.
		
\begin{figure}[ht!]
				\includegraphics[width=7cm]{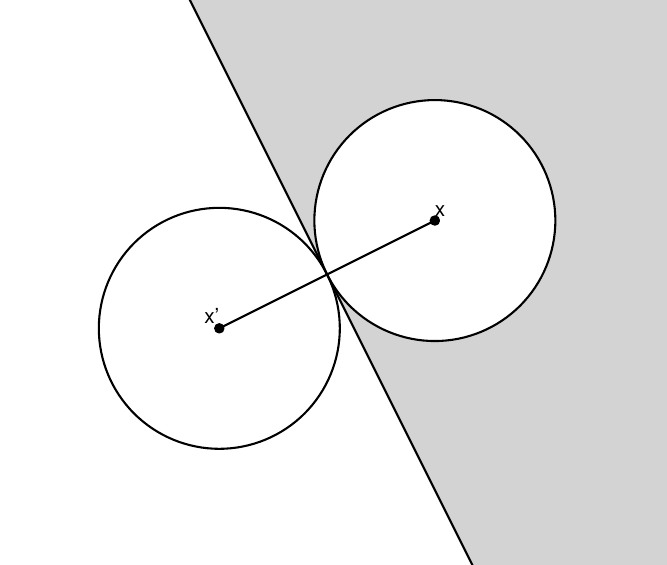}
				\caption{The set $M(x,x')$ from \eqref{pim10} is in gray}
				\label{fig.1}
				\end{figure}

		For $x,x'\in \R^2_+$ we define the set (see Figure \ref{fig.1}),
		 \begin{equation}\label{pim10}
			M(x,x')=\left\{y\in \R^2 : \frac{|x-x'|}{2}\leq|y-x|\leq |y-x'|\right\}.
			\end{equation}
			
		For all $y\in M(x,x')$, we have
		\[|x-y|^{-1}|x'-y|^{-1}\leq |x-y|^{-2}\,,\]
		so that for
		\[ \overline{B}_x(\delta) \equiv \left\{ y\ :\ |x-y| \leq \delta\right\}, \] 
		we have 
		\[\begin{split}&\int_{M(x,x')\cap \R_+^2}|x-y|^{-1}|x'-y|^{-1}e^{-\epsilon\lambda(|x-y|+|x'-y|)}\mathrm{d}y\\
		&\leq  \int_{\R_+^2\setminus \overline{B}_x\left(\frac{|x-x'|}{2}\right)}|x-y|^{-2}e^{-\epsilon\lambda|x-y|}\mathrm{d}y\\
		&\leq 2\pi   \int_{\frac{|x-x'|}{2}}^{+\infty}r^{-2}e^{-\epsilon\lambda r}r\mathrm{d}r\leq C\ln(2+|x-x'|^{-1})\,,
		\end{split}\]
uniformly in $\lambda\geq 1$.
		Then,
		
		\[\begin{split}&\int_{\overline{B}_x\left(\frac{|x-x'|}{2}\right)\cap \R_+^2}|x-y|^{-1}|x'-y|^{-1} e^{-\epsilon\lambda(|x-y|+|x'-y|)}\mathrm{d}y\\ 
		&\leq 2|x'-x|^{-1} \int_{\overline{B}_x\left(\frac{|x-x'|}{2}\right)}|x-y|^{-1} e^{-\epsilon\lambda|x-y|}\mathrm{d}y\\
		&= 4\pi |x'-x|^{-1} \int_{0}^{\frac{|x-x'|}{2}}r^{-1}e^{-\epsilon \lambda r}r\mathrm{d}r\leq 2\pi \,.
		\end{split}\]
This establishes the desired estimate on $\R_+^2 \cap \left( M(x,x') \cup \overline{B}_x\left( \frac{|x-x'|}{2} \right) \right)$. On the complement of this set in $\R_+^2$, the result follows simply by interchanging the roles of $x$ and $x'$.

    \vspace{0.2cm}

	\noindent\textbf{Off diagonal continuity:} 
	We write	\[\mathbb{A}(x,x')=\int_{\R^2_+}\frac{|x-y|K_{0, \lambda}(x,y)}{|x-y|}W(y)\frac{|x'-y|K_{0,\lambda}(y,x')}{|x'-y|}\mathrm{d}y\,.\]
	Let us consider $x_0\neq x'_0$. Denoting $\overline{B}_{z}\equiv \overline{B}_{z}(|x_0-x'_0|/4)\cap \R_+^2$ with $z\in \R_+^2$,  for all $(x,x')\in \overline{B}_{x_0}\times \overline{B}_{x'_0}$ and all $y\in \R_+^2\setminus (\overline{B}_{x_0}\cup \overline{B}_{x'_0})$, we have,

\begin{equation*}\label{eq:dist-xx}
|x - x'| \geq \frac{1}{2}|x_0 - x'_0|,
\end{equation*}
and 
\[
\frac{1}{2}|x_0 - x'_0| \leq |x - x'| \leq |y - x| + |y - x'| \leq 2 \max \big( |y - x|, |y - x'| \big),
\]
so that
\begin{equation*}\label{eq:min-bound}
\min \left( \frac{1}{|y - x|}, \frac{1}{|y - x'|} \right) = \frac{1}{\max \big( |y - x|, |y - x'| \big)} \leq \frac{4}{|x_0 - x'_0|},
\end{equation*}
and
\begin{align*}
\frac{1}{|y-x|} \frac{1}{|y-x'|} &= \min \left( \frac{1}{|y - x|}, \frac{1}{|y - x'|} \right) \max \left( \frac{1}{|y - x|}, \frac{1}{|y - x'|} \right) \\
&\leq \frac{4}{|x_0 - x'_0|} \max \left( \frac{1}{|y - x|}, \frac{1}{|y - x'|} \right).
\end{align*}
Then, by Proposition \ref{prop.free},
\[\begin{split}
	&\frac{|x-y||K_{0, \lambda}(x,y)|}{|x-y|}|W(y)|\frac{|x'-y||K_{0, \lambda}(y,x')|}{|x'-y|}\\
&\leq \frac{4|x-y||K_{0, \lambda}(x,y)|}{|x_0-x'_0|}|W(y)|\, |x'-y|\, |K_{0, \lambda}(y,x')|\, \max\left(|y-x|^{-1}\, ,\, |y-x'|^{-1}\right)\\
&\leq C\, e^{-\epsilon \, |y|}\, \max\left(|y-x|^{-1}\, ,\, |y-x'|^{-1}\right)
\,.\end{split}\]
If both $|x-x_0|, |x'-x_0'|\leq |x_0-x_0'|/8$ then the above function is dominated by $C e^{-\epsilon |y|}$, hence by dominated convergence, we get that $$(x,x')\mapsto \int_{\R_+^2\setminus (\overline{B}_{x_0}\cup \overline{B}_{x'_0})}K_{0, \lambda}(x,y)W(y)K_{0, \lambda}(y,x')\mathrm{d}y$$
is continuous at $(x_0,x'_0)$.

Now we analyze the integral restricted to $y\in \overline{B}_{x_0}$. { Fix  $0<\epsilon< 1$ and consider the function $F_\epsilon(x,x',y):=|x-y|^{1+\epsilon}K_{0,\lambda}(x,y)K_{0,\lambda}(y,x')$ which is jointly continuous in $x\in \overline{B}_{x_0}$ and $x'\in \overline{B}_{x_0'}$ at fixed $y$. We want to show that the map 

\[ (x,x')\mapsto \int_{\overline{B}_{x_0}}\frac{F_\epsilon(x,x',y)}{|x-y|^{1+\epsilon}}W(y)\mathrm{d}y\]
is continuous at $(x_0,x_0')$. For all $0<\delta\ll |x_0-x_0'|/8$ we have 
\begin{align*}
   & \Big | \int_{\overline{B}_{x_0}}\frac{F_\epsilon(x,x',y)}{|x-y|^{1+\epsilon}}W(y)\mathrm{d}y-\int_{\overline{B}_{x_0}}\frac{F_\epsilon(x_0,x_0',y)}{|x_0-y|^{1+\epsilon}}W(y)\mathrm{d}y\Big |\\
   & \leq \int_{\overline{B}_{x_0}(\delta)\cap \R^2_+}\Big (\frac{|F_\epsilon(x,x',y) W(y)|}{|x-y|^{1+\epsilon}}+\frac{|F_\epsilon(x_0,x_0',y) W(y)|}{|x_0-y|^{1+\epsilon}}\Big )\mathrm{d}y \\
   &+\Big | \int_{\overline{B}_{x_0}\setminus \overline{B}_{x_0}(\delta)}\frac{F_\epsilon(x,x',y)}{|x-y|^{1+\epsilon}}W(y)\mathrm{d}y-\int_{\overline{B}_{x_0}\setminus \overline{B}_{x_0}(\delta)}\frac{F_\epsilon(x_0,x_0',y)}{|x_0-y|^{1+\epsilon}}W(y)\mathrm{d}y\Big |\\
   &\leq C\, \delta^{1-\epsilon} +\Big | \int_{\overline{B}_{x_0}\setminus \overline{B}_{x_0}(\delta)}\frac{F_\epsilon(x,x',y)}{|x-y|^{1+\epsilon}}W(y)\mathrm{d}y-\int_{\overline{B}_{x_0}\setminus \overline{B}_{x_0}(\delta)}\frac{F_\epsilon(x_0,x_0',y)}{|x_0-y|^{1+\epsilon}}W(y)\mathrm{d}y\Big |,
\end{align*}
where the first terms are controlled by Proposition \ref{prop.free}. 
For a fixed $\delta$, the difference of the integrals in the last line above converges to zero when $(x,x')$ goes to $(x_0,x_0')$ due to Lebesgue's dominated convergence theorem. Thus 
\[ \limsup_{(x,x')\to (x_0,x_0')}\Big | \int_{\overline{B}_{x_0}}\frac{F_\epsilon(x,x',y)}{|x-y|^{1+\epsilon}}W(y)\mathrm{d}y-\int_{\overline{B}_{x_0}}\frac{F_\epsilon(x_0,x_0',y)}{|x_0-y|^{1+\epsilon}}W(y)\mathrm{d}y\Big |\leq C \delta^{1-\epsilon}\]
for all $0<\delta\ll |x_0-x_0'|/8$, which means that the limit exists and equals zero. We proceed in the same way for the integral restricted to $y\in \overline{B}_{x'_0}$.
}

\qed



\subsection{Proof of Lemma \ref{lem.A2}}
We have
\[AW(\mathscr{D}_0-i\lambda)^{-1}(x,x')=\int_{\R^2_+}\mathbb{A}(x,y)W(y)K_{0, \lambda}(y,x')\mathrm{d}y\,.\]		
\noindent\textbf{Proof of the bound in \eqref{pim12}.} By applying \eqref{pim11} we estimate 
\[\begin{split}|AW(\mathscr{D}_0-i\lambda)^{-1}(x,x')|&\leq \|W\|_\infty\int_{\R^2_+}|\mathbb{A}(x,y)||K_{0, \lambda}(y,x')|\mathrm{d}y\\
&\leq Ce^{-(c-\epsilon)\lambda|x-x'|} \int_{\R^2_+}\ln(2+|x-y|^{-1})|x'-y|^{-1}e^{-\epsilon\lambda |x-y|-\epsilon \lambda |x'-y|}\mathrm{d}y \,.\end{split}\]	
We have (see \eqref{pim10} for the definition of $M(x,x')$)
\begin{equation*}
\begin{split}&\int_{M(x,x')}\ln(2+|x-y|^{-1})|x'-y|^{-1}e^{-\epsilon\lambda|x-y|-\epsilon\lambda|x'-y|}\mathrm{d}y\\
&\leq\int_{M(x,x')}\ln(2+|x-y|^{-1})|x-y|^{-1}e^{-\epsilon\lambda|x-y|}\mathrm{d}y\\
&\leq 2\pi\int_{|x-x'|/2}^{+\infty}\ln(2+r^{-1})e^{-\epsilon\lambda r}\mathrm{d}r\leq C\,.\end{split}
\end{equation*}
Similarly, by interchanging $x$ with $x'$ we have
\begin{equation*}
	\begin{split}&\int_{M(x',x)}\ln(2+|x-y|^{-1})|x'-y|^{-1}e^{-\epsilon \lambda |x-y|-\epsilon \lambda |x'-y|}\mathrm{d}y\\
		&\leq\int_{M(x',x)}\ln(2+|x'-y|^{-1})|x'-y|^{-1}e^{-\epsilon\lambda|x'-y|}\mathrm{d}y\\
		&\leq2\pi\int_{|x-x'|/2}^{+\infty}\ln(2+r^{-1})e^{-\epsilon\lambda r}\mathrm{d}r\leq C\,.\end{split}
\end{equation*}
We also have
\begin{equation*}
	\begin{split}&\int_{\overline{B}_x(|x-x'|/2)}\ln(2+|x-y|^{-1})|x'-y|^{-1}e^{-\epsilon\lambda|x-y|-\epsilon\lambda|x'-y|}\mathrm{d}y\\
		&\leq 2|x'-x|^{-1}\int_{\overline{B}_x(|x-x'|/2)}\ln(2+|x-y|^{-1})e^{-\epsilon\lambda|x-y|}\mathrm{d}y\\ 
		&= 4\pi |x'-x|^{-1}\int_{0}^{|x-x'|/2}r\ln(2+r^{-1})e^{-\epsilon\lambda r}\mathrm{d}r\leq C\,.
		\end{split}
	\end{equation*}
Similarly,
\begin{equation*}
	\begin{split}&\int_{\overline{B}_{x'}(|x-x'|/2)}\ln(2+|x-y|^{-1})|x'-y|^{-1}e^{-\epsilon\lambda|x-y|-\epsilon\lambda|x'-y|}\mathrm{d}y\\
		&\leq \ln(2+|x'-x|^{-1})\int_{\overline{B}_{x'}(|x-x'|/2)}|x'-y|^{-1}e^{-\epsilon\lambda|x'-y|}\mathrm{d}y\leq  C\,.		
	\end{split}
\end{equation*}

\noindent \textbf{Proof of continuity.}  The continuity at $(x_0,x'_0)$ for $x_0\neq x'_0$ can be established as in the proof of Lemma \ref{lem.A}.

 Now let us consider $x_0=x'_0$. Let $\delta\in(0,1)$ and take $x,x'\in \overline{B}_{x_0}(\delta/4)\cap \R^2_+$. We write, for any $\epsilon\in(0,\frac12)$,
\[AW(\mathscr{D}_0-i\lambda)^{-1}(x,x')=\int_{\R^2_+}\frac{G_\epsilon(x,x',y)}{|x-y|^\epsilon|x'-y|^{1+\epsilon}}\mathrm{d}y\,,\]
where $G_\epsilon$ is the following continuous function {on $x$ and $x'$ at fixed $y$:}
\[G_\epsilon(x,x',y)=|x-y|^{\epsilon}\mathbb{A}(x,y)W(y)|x'-y|^{1+\epsilon}K_{0, \lambda}(y,x')\,.\]		
 We write
\begin{align*}\int_{\R^2_+}\frac{G_\epsilon(x,x',y)}{|x-y|^\epsilon|x'-y|^{1+\epsilon}}\mathrm{d}y&=\int_{\overline{B}_{x_0}(1)\cap \R^2_+}\frac{G_\epsilon(x,x',y)}{|x-y|^\epsilon|x'-y|^{1+\epsilon}}\mathrm{d}y\\
&\qquad +\int_{\R^2_+\setminus \overline{B}_{x_0}(1)}\frac{G_\epsilon(x,x',y)}{|x-y|^\epsilon|x'-y|^{1+\epsilon}}\mathrm{d}y\, .
\end{align*}
The second term on the right hand side is continuous because the integrand is dominated by an $L^1$ function and has no singularities. 

We now analyze the first integral for $x_0=x_0'$. We have 
\begin{equation*}\begin{split}
&		\left \vert \int_{\overline{B}_{x_0}(1)\cap \R^2_+}\frac{G_\epsilon(x,x',y)}{|x-y|^\epsilon|x'-y|^{1+\epsilon}}\mathrm{d}y-\int_{\overline{B}_{x_0}(1)\cap \R^2_+}\frac{G_\epsilon(x_0,x_0,y)}{|x_0-y|^{1+2\epsilon}}\mathrm{d}y \right \vert \\
&\quad \qquad\leq \int_{\overline{B}_{x_0}(\delta)\cap \R^2_+}\left ( \frac{|G_\epsilon(x,x',y)|}{|x-y|^\epsilon|x'-y|^{1+\epsilon}}+\frac{|G_\epsilon(x_0,x_0,y)|}{|x_0-y|^{1+2\epsilon}}\right ) \, \mathrm{d}y\\
&\qquad \qquad +\int_{(\overline{B}_{x_0}(1)\setminus \overline{B}_{x_0}(\delta))\cap \R_+^2}\left \vert\frac{G_\epsilon(x,x',y)}{|x-y|^\epsilon|x'-y|^{1+\epsilon}}-\frac{G_\epsilon(x_0,x_0,y)}{|x_0-y|^{1+2\epsilon}}\right \vert \, \mathrm{d}y \, .
\end{split}\end{equation*}
We take the $\limsup$ when $(x,x')\to (x_0,x_0)$ on both sides of the inequality. The second term on the right hand side converges to zero due to Fatou's lemma and the continuity of the integrand. For the first term we observe that by splitting the integral according to $|x-y|\leq |x'-y|$ and $|x-y|>|x'-y|$, we get
\[\int_{\overline{B}_{x_0}(\delta)\cap \R^2_+}\frac{|G_\epsilon(x,x',y)|}{|x-y|^\epsilon|x'-y|^{1+\epsilon}}\mathrm{d}y\leq C\delta^{1-2\epsilon}\,\]
hence its $\limsup$ will obey the same inequality. Since $0< \epsilon <1/2 $, and because $0<\delta<1$ is arbitrary, we obtain the desired continuity at $x_0$. \qed

\section{The resolvent kernel for \texorpdfstring{$W\neq 0$}{W != 0} and \texorpdfstring{$b\neq 0$}{b != 0}}\label{resolventmagnetic}

So far, we treated the case without magnetic field. In this section,  we turn it on. In view of the symmetries of the system (see \cite[Remark 1.10]{MR4769230}) it is enough to consider the case $b>0$. 

We recall that the vector potential $\mathbf{A}$ is given in \eqref{eq.ax_2} and can be written as $\mathbf{A}=ba$ with $a(x)=(-x_2,0)$. The operator we are interested in can be written as 
\[\mathscr{D}_{b,W}=\mathscr{D}_0+W-b\sigma\cdot a(x)\,.\]
We define 
\begin{equation}\label{tigre1}
\varphi(x,x')=(x_1'-x_1)\frac{x_2+x_2'}{2}\,,
\end{equation}
so that,
$\nabla_x\varphi(x,x')=-2^{-1}(x_2+x_2',x_1-x_1')$ and 
$$a(x)-\nabla_x \varphi(x,x')=2^{-1}(-x_2+x_2',x_1-x_1')=a_t(x-x'),$$
where $a_t(x):=2^{-1}(-x_2,x_1)$.
We have the following important local gauge transform identity: 
\begin{equation}\label{pim14} \mathscr{D}_{b,W}e^{ib\varphi(\cdot ,x')}=e^{ib\varphi(\cdot ,x')}\big (\mathscr{D}_{0,W}-b\sigma\cdot a_t(\cdot -x')\big )\,.
\end{equation}

Let us consider the following integral kernels:
\begin{equation}\label{tigre4}\begin{split}
S_b(x,x';\lambda)&=e^{ib\varphi(x,x')}(\mathscr{D}_{0,W}-i\lambda)^{-1}(x,x')\,,\\
T_b(x,x';\lambda)&=b\sigma\cdot  a_t(x-x')e^{ib\varphi(x,x')}(\mathscr{D}_{0,W}-i\lambda)^{-1}(x,x')\,.\end{split}
\end{equation}
By using the Schur test and the bound of Proposition \ref{prop.resolventWz}, we get, for all $\lambda\geq 1$,
\begin{equation}\label{eq.Tb}
\|T_b(\lambda)\|\leq Cb|\lambda|^{-2}\,.
\end{equation}
In the next lemma we show that  $S_b(\lambda)$ is an approximated right inverse for $\mathscr{D}_{b,W}-i\lambda$.
\begin{lemma}\label{lem:parametrix}
Let $\lambda\neq 0$. We have $\mathrm{Ran} \, S_b(\lambda)\subset\mathrm{Dom}(\mathscr{D}_{b,W})$ and
	\[(\mathscr{D}_{b,W}-i\lambda)S_b(\lambda)=1-T_b(\lambda)\,.\]
	\end{lemma}
\begin{proof}  To simplify notation we write 
$$K_{W,\lambda}(x,x')\equiv (\mathscr{D}_{0,W}-i\lambda)^{-1}(x,x').$$

    Let $0\leq \psi\leq 1$ with $\psi\in C_0^\infty(\R^2)$ such that 
$\sum_{\ell\in \mathbb{Z}^2} \psi(x-\ell)=1$ for all $x\in \R^2$. Denote by $\psi_{\ell,\epsilon}(x)=\psi(x\,\epsilon^{-1}-\ell)$. We then have $\sum_{\ell\in \mathbb{Z}^2} \psi_{\ell,\epsilon}(x)=1$ for all $x\in \R^2$, and the support of $\psi_{\ell,\epsilon}$ has a diameter of order $\epsilon$. 

Let us consider $f\in C_0^\infty(\R^2_+,\C^2)$. We will show that $\sum_{\ell\in \mathbb{Z}^2}e^{ib\varphi(\cdot,\ell\, \epsilon)}(\mathscr{D}_{0,W}-i\lambda)^{-1}\psi_{\ell,\epsilon} \, f$ converges strongly to $S_b(\lambda)f$. 
We have the estimate 
\begin{equation}\label{tigre7}
\begin{split}
|e^{ib\varphi(x,\ell\, \epsilon)}-e^{ib\varphi(x,y)}|&\leq \, 2\,  \sin |b\varphi(x,\ell\, \epsilon -y)/2|\\
&\leq b |\varphi(x,\ell\, \epsilon -y)|\leq (b/2)\,  (|x-y| +|y|)\,|\ell\, \epsilon -y|.
\end{split}
\end{equation}

We use this inequality in order to get a pointwise estimate, for all $x\in \R_+^2$, 
\begin{equation}\label{tigre8}
\begin{aligned}
 &\Big |\sum_{\ell\in \mathbb{Z}^2}e^{ib\varphi(x,\ell\, \epsilon)}\big ((\mathscr{D}_{0,W}-i\lambda)^{-1}\psi_{\ell,\epsilon} \, f \big )(x)- \big (S_b(\lambda)f \big )(x)\Big | \\
 &\leq \sum_{\ell\in \mathbb{Z}^2} \int_{\R^2_+}  |K_{W,\lambda}(x,y)|\, \psi_{\ell,\epsilon}(y)\, |f(y)|\, |e^{ib\varphi(x,\ell\, \epsilon)}-e^{ib\varphi(x,y)}|\, \mathrm{d} y\\
 &\leq (b/2)\sum_{\ell\in \mathbb{Z}^2} \int_{\R^2_+}  |K_{W,\lambda}(x,y)|\, \psi_{\ell,\epsilon}(y)\, |f(y)|\, (|x-y| +|y|)\,|\ell\, \epsilon -y| \, \mathrm{d} y\\
 &\leq \epsilon \, C\, \sum_{\ell\in \mathbb{Z}^2} \int_{\R^2_+}  |K_{W,\lambda}(x,y)|\, \psi_{\ell,\epsilon}(y)\, |f(y)|\, (|x-y| +|y|) \, \mathrm{d} y\\
 &=\epsilon \, C \, \int_{\R^2_+}  |K_{W,\lambda}(x,y)|\,|x-y|\, |f(y)|\,\mathrm{d} y + \epsilon \, C \, \int_{\R^2_+}  |K_{W,\lambda}(x,y)|\,|y|\, |f(y)|\,\mathrm{d} y. 
\end{aligned}
\end{equation}
By taking the $L^2$ norm of the latter two terms, and using that $|K_{W,\lambda}(x,y)|$ and $|x-y|\, |K_{W,\lambda}(x,y)|$ are bounded by $L^1$ functions of $x-y$, together with the fact that $f$ has compact support, we get the claimed strong convergence when $\epsilon\to 0$ by Young's inequality. 

Let $g\in {\rm Dom}\big (\mathscr{D}_{b,W}\big )$. We have, using \eqref{pim14}
\begin{equation*}\begin{aligned}
&\langle S_b(\lambda)f,(\mathscr{D}_{b,W}+i\lambda) g\rangle= \lim_{\epsilon\to 0}\sum_{\ell\in \mathbb{Z}^2} \langle e^{ib\varphi(\cdot,\ell\, \epsilon)}(\mathscr{D}_{0,W}-i\lambda)^{-1}\psi_{\ell,\epsilon} \, f\,,\, (\mathscr{D}_{b,W}+i\lambda) g\rangle \\
&\qquad =  \lim_{\epsilon\to 0}\sum_{\ell\in \mathbb{Z}^2} \langle (\mathscr{D}_{0,W}-i\lambda)^{-1}\psi_{\ell,\epsilon} \, f\,,\, \big (\mathscr{D}_{0,W}+i\lambda-b\sigma\cdot a_t(\cdot -\ell\, \epsilon)\,\big ) e^{-ib\varphi(\cdot,\ell\, \epsilon)}g\rangle\, .
\end{aligned}
\end{equation*}
This yields
\begin{equation*}\begin{aligned}
&\langle S_b(\lambda)f,(\mathscr{D}_{b,W}+i\lambda) g\rangle= \lim_{\epsilon\to 0}\sum_{\ell\in \mathbb{Z}^2} \langle \psi_{\ell,\epsilon} \, f\,,\, e^{-ib\varphi(\cdot,\ell\, \epsilon)}g\rangle \\
&\qquad -b \lim_{\epsilon\to 0}\sum_{\ell\in \mathbb{Z}^2} \langle \sigma\cdot a_t(\cdot -\ell\, \epsilon)\, e^{ib\varphi(\cdot,\ell\, \epsilon)}\, (\mathscr{D}_{0,W}-i\lambda)^{-1}\psi_{\ell,\epsilon} \, f\,,\, g\rangle\, .
\end{aligned}
\end{equation*}
The latter two limits can be computed in the same fashion as in \eqref{tigre8} using the estimate in \eqref{tigre7}. Hence,
\begin{equation*}
	\big \langle S_b(\lambda)f,(\mathscr{D}_{b,W}+i\lambda) g\big \rangle
	=\big \langle f,g\rangle-\langle T_b(\lambda)f,g\big \rangle=\big \langle\big (1-T_b(\lambda)\big )f,g\big \rangle\,.
\end{equation*}
Since $S_b(\lambda)$ and $T_b(\lambda)$ are bounded operators, this equality can be extended to all $f\in L^2(\R^2_+,\C^2)$. This shows that $S_b(\lambda) f\in \mathrm{Dom}(\mathscr{D}_{b,W}+i\lambda)^*=\mathrm{Dom}(\mathscr{D}_b-i\lambda)$ and that
\[\langle(\mathscr{D}_{b,W}-i\lambda)S_b(\lambda)f,g\rangle=\langle \big (1-T_b(\lambda)\big )f,g\rangle\,.\]
The conclusion follows.
	\end{proof}
    
\subsection{The integral kernel of $(\mathscr{D}_{b,W}-i\lambda)^{-1}$ for large $\lambda$}
    
Thanks to \eqref{eq.Tb}, for all $|\lambda|$ large enough, $1-T_b(\lambda)$ is bijective and we have from Lemma~\ref{lem:parametrix}
\begin{equation}\label{tigre2}(\mathscr{D}_{b,W}-i\lambda)^{-1}=S_b(i\lambda)(1-T_b(i\lambda))^{-1}=\sum_{k=0}^1 S_b(i\lambda)T_b(i\lambda)^k+(\mathscr{D}_{b,W}-i\lambda)^{-1}T_b(i\lambda)^2\,.
\end{equation}
By taking the adjoint and change $\lambda$ with $-\lambda$, we get
\begin{equation}\label{tigre3}(\mathscr{D}_{b,W}-i\lambda)^{-1}=\sum_{k=0}^1 (T_b(-i\lambda)^*)^kS_b(-i\lambda)^*+(T_b(-i\lambda)^*)^2(\mathscr{D}_{b,W}-i\lambda)^{-1}\,.
\end{equation}
On the right hand side of \eqref{tigre2} we replace $(\mathscr{D}_{b,W}-i\lambda)^{-1}$ with the expression from \eqref{tigre3} and we obtain
\begin{equation}\label{eq.Db-il}
\begin{aligned}
(\mathscr{D}_{b,W}-i\lambda)^{-1}&=\sum_{k=0}^1 S_b(i\lambda)T_b(i\lambda)^k +\sum_{k=0}^1 (T_b(-i\lambda)^*)^kS_b(-i\lambda)^*T_b(i\lambda)^2\,\\
&\qquad +(T_b(-i\lambda)^*)^2(\mathscr{D}_{b,W}-i\lambda)^{-1}T_b(i\lambda)^2\,.
\end{aligned}
\end{equation}
On the right hand side of \eqref{eq.Db-il} we have several terms involving $S_b(\pm i\lambda)$ and powers of $T_b(\pm i\lambda)$. The integral kernel of $S_b(\pm i\lambda)$ from \eqref{tigre4} has the same behavior as the one of $(\mathscr{D}_{0,W} \mp i\lambda)^{-1}$, both with respect to its exponential localization near the diagonal and the local singularity.  On the other hand, the integral kernel of $T_b(\pm i\lambda)$ from \eqref{tigre4} is globally bounded due to the linear factor $a_t(x-x')$ which cancels the singularity of $(\mathscr{D}_{0,W} \mp i\lambda)^{-1}(x,x')$. Expanding as in \eqref{pim2} and reasoning as in Lemma \ref{lem.A2} we obtain that the integral kernel of $S_b(i\lambda)T_b(i\lambda)$ and its adjoint will be continuous and exponentially localized, and the same is true for $T_b(\pm i\lambda)^2$ and its adjoint. Note that the current setting is simpler : the integral kernel of $T_b(i\lambda)$ is bounded, whereas the kernel of $A$ in Lemma \ref{lem.A2} possesses logarithmic singularities.

The last term on the right hand side of \eqref{eq.Db-il} is treated by using Lemmas \ref{lem.Simon}, \ref{lem.CT}, \ref{lemma-pim} and Remark \ref{remark-pim}, as we did in the proof of Proposition \ref{prop.resolventWz}. In particular, this last term has a continuous integral kernel which is exponentially localized around the diagonal. 

We infer that, for some $\lambda_0\geq 1$, the resolvent $(\mathscr{D}_{b,W}-i\lambda)^{-1}$ has an integral kernel for all $|\lambda|\geq \lambda_0$, and the kernel of $(\mathscr{D}_{b,W}-i\lambda)^{-1}-S_b(i\lambda)$ is continuous and exponentially localized near the diagonal. Moreover, 
\begin{equation}\label{tigre5}|(\mathscr{D}_{b,W}-i\lambda)^{-1}(x,x')|\leq\frac{C}{|x-x'|}e^{-c\, |\lambda||x-x'|}\,,\quad \forall |\lambda|\geq \lambda_0\geq 1\, 
\end{equation}

\subsection{The integral kernel of $(\mathscr{D}_{b,W}-z)^{-1}$ for $z\notin \R$} Next, we want to study the resolvent at $z=\mu+i\lambda$ with $\mu\in\R$ and $\lambda\in \mathbb{R}\setminus\{0\}$.

We first study the case when $0<|\lambda|\leq \lambda_0$ with $\lambda_0$ from \eqref{tigre5}. 
To do so, we use the iterated first resolvent formula:
\begin{equation}\label{eq.reszil}
\begin{split}
(\mathscr{D}_{b,W}-z)^{-1}&=\sum_{j=0}^5(z-i\lambda_0)^j(\mathscr{D}_{b,W}-i\lambda_0)^{-j-1}\\
&\qquad +(z-i\lambda_0)^{6}(\mathscr{D}_{b,W}-i\lambda_0)^{-3}(\mathscr{D}_{b,W}-z)^{-1}(\mathscr{D}_{b,W}-i\lambda_0)^{-3}\,.
\end{split}
\end{equation}
Arguing as in the proof of Proposition \ref{prop.resolventWz} and as in the previous subsection we get 
\[|(\mathscr{D}_{b,W}-z)^{-1}(x,x')|\leq\, C\max\left( |\lambda|^{-1},1\right)\, \frac{\langle\mu\rangle^6}{|x-x'|}e^{-c|\lambda||x-x'|}\,,\]

It remains to investigate the resolvent for $|\lambda|\geq \lambda_0$ and $\mu\in \R$. The existence of an integral kernel  follows again from \eqref{eq.reszil} with $z=\mu+i\lambda$ and $i\lambda_0$  replaced by $i\lambda$, and using \eqref{tigre5}. 
This ends the proof of Theorem \ref{thm.main}(i) and (ii).

\section{Functions of edge and bulk operators}
\label{sec:functions}

Let $F\in \mathscr{S}(\R)$ be any real-valued function in the Schwartz class.
Let us recall the Helffer--Sj\"{o}strand formula \cite{davies1995functional}. We  consider an almost analytic extension $\widetilde{F}_N$ of $F$ constructed as follows. 

Let $0\leq \chi\leq 1$ be a smooth function on $\mathbb{R}$ such that 
$ \chi(v)=\left \{ \begin{matrix}
    1, & \text{ if }|v|\leq 1/2\\
    0, & \text{ if }|v|\geq 1
\end{matrix}
\right . \, .$
Then we define
\begin{equation}\label{tigre10}
\widetilde{F}_N{(u,v)}:=\sum_{j=0}^N \frac{F^{(j)}(u)}{j!}\, (i v)^j\, \chi(v),\quad N\geq 2.
\end{equation}
We have $$(\partial_u +i\partial_v)\widetilde{F}_N{(u,v)}=\frac{F^{(N+1)}(u)}{N!}\, (i v)^N\, \chi(v)+i\sum_{j=0}^N \frac{F^{(j)}(u)}{j!}\, (i v)^j\, \chi'(v)\, ,$$
which means that $\overline{\partial} \widetilde{F}_N$ decays faster than any power of $\langle u\rangle$ and vanishes as $|v|^N$ when $v\to 0$. We then have 
\begin{equation}\label{tigre11}
F(\mathscr{D}_{b,W})=-\pi^{-1} \int_{\R\times [-1,1]}\, \overline{\partial} \widetilde{F}_N(u,v)\, \big (\mathscr{D}_{b,W}-(u+iv)\big )^{-1}\, \mathrm{d}u\, \mathrm{d}v\,, \quad \forall N\geq 2. 
\end{equation}

\subsection{Proof of Theorem \ref{thm-tigre}(i)} 
Using \eqref{eq.reszil} and noticing that all the analytic terms will vanish after integration, we have (we put $z=u+iv$) for $N\geq 2$:
\begin{equation}\label{tigre12}
\begin{split}
F(\mathscr{D}_{b,W}) = & -\pi^{-1} \int_{\mathbb{R}\times [-1,1]}\, \overline{\partial} \widetilde{F}_N(u,v)\, (z-i\lambda_0)^6 \\
& \times (\mathscr{D}_{b,W}-i\lambda_0)^{-3}\, \big (\mathscr{D}_{b,W}-z\big )^{-1}\, (\mathscr{D}_{b,W}-i\lambda_0)^{-3}\, \mathrm{d}u\, \mathrm{d}v\,.
\end{split}
\end{equation}
The polynomial growth in $u$ is controlled by the fact that $F$ is a Schwartz function. 

The operator in the middle has an integral kernel given by
$$M_{jk}(x,x';z)=\big \langle (\mathscr{D}_{b,W}+i\lambda_0)^{-3}(\cdot ,x)e_j\,,\big (\mathscr{D}_{b,W}-z\big )^{-1}\, (\mathscr{D}_{b,W}-i\lambda_0)^{-3}(\cdot,x')e_k\big \rangle_{L^2(\R_+^2,\C^2)}\, .$$
Applying as in Lemma \ref{lemma-pim} with $A_1^*=A_2=(\mathscr{D}_{b,W}-i\lambda_0)^{-3}$, there exist two constants $C_0,c>0$ such that for every $0<|v|\leq 1$ and for all $x,x'\in \R^2_+$ we have 
$$\big |  M_{jk}(x,x';z)\big | \leq C_0\, |v|^{-1} e^{-c\, |v|\, |x-x'|}\, .$$
Moreover, this integral kernel is holomorphic as a function of $z$ on $\C\setminus \R$ because the resolvent in the middle is holomorphic.  In addition, for every $m\in \mathbb{N}$ we have 
$$\langle x-x'\rangle^m \big |  M_{jk}(x,x';z)\big | \leq \tilde{C}\, |v|^{-m-1} $$
where we traded off the exponential decay with a polynomial one, the price being a more singular behavior in $1/|v|$. But any such polynomial singularity in $1/|v|$ can be controlled by choosing any $N\geq m+1$ when we construct the almost analytic extension $\widetilde{F}_N$. 

Regarding the bulk operator, from Remark \ref{remark.tigre} we know that all the estimates from Theorem \ref{thm.main} can be extended to the bulk operator. This proves that the above desired properties hold true for $F(\mathrm{D}_{b,W})$ as well. \qed 

\subsection{Proof of Theorem \ref{thm-tigre}(ii)} 

Let $0\leq \eta(x_2) \leq 1$ be a smooth cut-off function which equals $1$ if $0\leq x_2\leq 1/2$  and equals $0$ if $1\leq x_2$. Let us define 
$$L(z)= \eta \, (\mathscr{D}_{b,W}-z)^{-1} +(1-\eta)\, (\mathrm{D}_{b,W}-z)^{-1}\, \chi_{\R^2_+} .$$
Since $1-\eta$ vanishes at the boundary, $ (1-\eta)\, (\mathrm{D}_{b,W}-z)^{-1}$ takes its values in the domain of $\mathscr{D}_{b,W}$ and 
$$(\mathscr{D}_{b,W}-z)\, L(z)=1+M(z) \text{ and } M(z)=-i\eta'\, \sigma_2 (\mathscr{D}_{b,W}-z)^{-1} +i\eta'\, \sigma_2 (\mathrm{D}_{b,W}-z)^{-1}\, \chi_{\R^2_+}.$$
This implies
$$(\mathscr{D}_{b,W}-z)^{-1}=L(z)-(\mathscr{D}_{b,W}-z)^{-1}\, M(z).$$
Applying it with \eqref{tigre11} on $\mathscr{D}_{b,W}$ and $\mathrm{D}_{b,W}$, we obtain
\begin{equation}\label{tigre13}
\begin{split}
F(\mathscr{D}_{b,W})&=\eta \,F(\mathscr{D}_{b,W}) +(1-\eta)\, F(\mathrm{D}_{b,W})\, \chi_{\R^2_+} \\
&\qquad +\pi^{-1} \int_{\R\times [-1,1]}\, \overline{\partial} \widetilde{F}_N(u,v)\, \big (\mathscr{D}_{b,W}-(u+iv)\big )^{-1}\, M(u+iv)\, \mathrm{d}u\, \mathrm{d}v\,.
\end{split}
\end{equation}
Let $0\leq \tilde{\eta}(x_2)\leq 1$ be another smooth cutoff function with support in $x_2>2$ and which equals $1$ if $x_2>3$. Then $\tilde{\eta}\, \eta'=0$ and hence $\tilde{\eta}\, M(z)=0$. We multiply \eqref{tigre13} with $\tilde{\eta}$ to the left and obtain 
\begin{equation}\label{tigre14}
\begin{split}
&\tilde{\eta}\, F(\mathscr{D}_{b,W})=\tilde{\eta}\, F(\mathrm{D}_{b,W})\, \chi_{\R^2_+} \\
& +\pi^{-1} \int_{\R\times [-1,1]}\, \overline{\partial} \widetilde{F}_N(u,v)\, \big [\tilde{\eta}\, ,\, \big (\mathscr{D}_{b,W}-(u+iv)\big )^{-1}\big ] \, M(u+iv)\, \mathrm{d}u\, \mathrm{d}v\, .
\end{split}
\end{equation}
We have 
\[\begin{split}
    \big [\tilde{\eta}\, ,\, \big (\mathscr{D}_{b,W}-z\big )^{-1}\big ] \, M(z)
&=\big (\mathscr{D}_{b,W}-z\big )^{-1}
 \big [ \big (\mathscr{D}_{b,W}-z\big )\, ,\,\tilde{\eta}\big ]
\big (\mathscr{D}_{b,W}-z\big )^{-1}\, M(z)
\\&=-i \,(\mathscr{D}_{b,W}-z\big )^{-1}\, \sigma_2\, \tilde{\eta}'\, (\mathscr{D}_{b,W}-z\big )^{-1}\, M(z).
\end{split}\]
The right hand side is the composition of three resolvents which has the following integral kernel
$$\mathfrak{S}(x,x';z)=-i\int_{\R_+^2\times \R_+^2}(\mathscr{D}_{b,W}-z\big )^{-1}(x,y)\, \sigma_2\, \tilde{\eta}'(y)\, (\mathscr{D}_{b,W}-z\big )^{-1}(y,y')\, M(y',x';z)\, \mathrm{d}y\, \mathrm{d}y'\,  .$$
Using the pointwise estimate from Theorem \ref{thm.main}(ii) and reasoning as done in the proof of Lemma~\ref{lem.A} we obtain the following estimate for some finite integer $p>1$
$$|\mathfrak{S}(x,x';z)|\leq C\, \frac{\langle u\rangle^p}{|v|^p}\, \int_{\R_+^2} \frac{e^{-c|v||x-y|}}{|x-y|}\, |\tilde{\eta}'(y)|\, \ln\big (2+|y-x'|^{-1}\big )e^{-c|v||y-x'|}\, \mathrm{d}y\, \,  .$$
Now reasoning as in the proof of Lemma \ref{lem.A2} we can show that the integral defining $\mathfrak{S}(x,x';z)$ is jointly continuous as a function of $x$ and $x'$. Moreover, because $\tilde{\eta}'(y)$ is supported in the region with $0\leq y_2\leq 3$ we have that for every $m\geq 1$
\begin{equation*}
\begin{split}
\langle x_2\rangle^m\, |\mathfrak{S}(x,x';z)|&\leq C_m\, \frac{\langle u\rangle^p}{|v|^{p+m}}\, \int_{\R_+^2} \frac{e^{-c\frac{|v|}{2}|x-y|}}{|x-y|}\, |\tilde{\eta}'(y)|\, \ln\big (2+|y-x'|^{-1}\big )e^{-c|v||y-x'|}\, \mathrm{d}y\, \\
&\leq \tilde{C}_m\, \frac{\langle u\rangle^{p}}{|v|^{{p'}+m}}
\end{split}
\end{equation*}
for some $p'\geq p$. Introducing this in \eqref{tigre14} and choosing $N>p'+m$ we obtain the result. \qed

\subsection*{Acknowledgments}
This work was partially conducted within the France 2030 framework programme, the Centre Henri Lebesgue  ANR-11-LABX-0020--01.  It has also been partially supported by CNRS International Research Project Spectral Analysis of Dirac Operators---SPEDO. H.C. acknowledges support from the Independent Research Fund Denmark–Natural Sciences, grant DFF-10.46540/2032-00005B, and from DNRF Centre CLASSIQUE sponsored by the Danish National Research Foundation. E.S.  acknowledges  support from  Fondecyt (ANID, Chile) 
through the  grants \# 123--1539 and \#  125--0596.

The authors are very grateful to
CIRM (and its staff) where this work was initiated and developed.  

		\appendix

        \section{Proof of the integral kernel formula \eqref{eq.KR}}
\label{app:proof_KR}

In this appendix, we provide the elementary proof of the formula \eqref{eq.KR} for the resolvent kernel of the one-dimensional free Dirac operator on the real line. 

Let us apply the Fourier transform $\mathscr{F}_{x_2\to k}$ to the equation $(\mathscr{D}^\R_{0}(\xi)-i)u = f$. Using the property that $\mathscr{F}_{x_2\to k}(\partial_2 u)(k) = i k \mathscr{F}_{x_2\to k}(u)(k)$, the operator $\mathscr{D}^\R_{0}(\xi)-i$ becomes a multiplication by the symbol matrix
\begin{equation*}
M(k) = \begin{pmatrix} -i & \xi - i k \\ \xi + i k & -i \end{pmatrix}\,.
\end{equation*}
The determinant of this matrix is $\det(M(k)) = (-i)^2 - (\xi - i k)(\xi + i k) = -(k^2 + \xi^2 + 1) = -(k^2 + \langle \xi \rangle^2)$. 
Consequently, its inverse in Fourier space is given by
\begin{equation*}
M(k)^{-1} = \frac{-1}{k^2 + \langle \xi \rangle^2} \begin{pmatrix} -i & -\xi + i k \\ -\xi - i k & -i \end{pmatrix} 
= \frac{i}{k^2 + \langle \xi \rangle^2} \begin{pmatrix} 1 & -k - i\xi \\ k - i\xi & 1 \end{pmatrix}\,.
\end{equation*}
The integral kernel $K^\R_{0,\xi}(x_2)$ is obtained by taking the inverse Fourier transform $\mathscr{F}^{-1}_{k\to x_2}$ of $M(k)^{-1}$. To compute this, we recall the following two standard inverse Fourier transforms for $\beta = \langle \xi \rangle > 0$:
\begin{equation*}
I_1(x_2) = \mathscr{F}^{-1}_{k\to x_2}\left(\frac{1}{k^2 + \beta^2}\right)(x_2) = \frac{e^{-\beta |x_2|}}{2\beta}\,,
\end{equation*}
and
\begin{equation*}
I_2(x_2) = \mathscr{F}^{-1}_{k\to x_2}\left(\frac{k}{k^2 + \beta^2}\right)(x_2) = -i \partial_{x_2} I_1(x_2) = i\, \sign(x_2) \frac{e^{-\beta |x_2|}}{2}\,.
\end{equation*}
Using these integrals, we can explicitly evaluate each entry of the matrix $K^\R_{0,\xi}(x_2)$:
\begin{itemize}
    \item For the diagonal entries, we simply have $i I_1(x_2) = \frac{i}{2\langle \xi \rangle} e^{-\langle \xi \rangle |x_2|}$.
    \item For the top-right entry, we obtain:
    \begin{equation*}
        -i I_2(x_2) + \xi I_1(x_2) = \frac{\sign(x_2)}{2} e^{-\langle \xi \rangle |x_2|} + \frac{\xi}{2\langle \xi \rangle} e^{-\langle \xi \rangle |x_2|} = \frac{e^{-\langle \xi \rangle |x_2|}}{2\langle \xi \rangle} \left( \langle \xi \rangle \sign(x_2) + \xi \right)\,.
    \end{equation*}
    \item For the bottom-left entry, we obtain:
    \begin{equation*}
        i I_2(x_2) + \xi I_1(x_2) = -\frac{\sign(x_2)}{2} e^{-\langle \xi \rangle |x_2|} + \frac{\xi}{2\langle \xi \rangle} e^{-\langle \xi \rangle |x_2|} = \frac{e^{-\langle \xi \rangle |x_2|}}{2\langle \xi \rangle} \left( -\langle \xi \rangle \sign(x_2) + \xi \right)\,.
    \end{equation*}
\end{itemize}
Combining these terms yields the exact matrix expression presented in \eqref{eq.KR}.

\section{Three technical lemmas}

{ The first lemma is an adaptation of \cite[Lemma B.7.8]{MR670130}}
\begin{lemma}[Existence of continuous kernels]\label{lem.Simon}
	Let $A_1$ and $A_2$ be two integral operators with kernels
    $$\R_+^2\times \R_+^2\ni (x,x')\mapsto \mathbb{A}_j(x,x')\in \C^{2\times 2},\quad j\in \{1,2\},$$
    such that the maps $x'\mapsto \mathbb{A}_j(\cdot,x')\in L^2(\mathbb{R}^2_+,\mathbb{C}^{2\times2})$ are continuous and uniformly bounded. Let $B$ be a bounded operator on $L^2(\R^2_+,\mathbb{C}^{2})$ and let $\{e_k\}_{k=1}^2$ be the canonical basis of $\C^2$. 
	
	Then, $A_1^* BA_2$ has a continuous integral kernel which is given, for $j,k\in \{1,2\}$, by
	\[\begin{split}\mathcal{K}_{kj}(x,x'):=\langle  e_k,(A_1^*BA_2)(x,x')e_j\rangle_{\mathbb{C}^2}=\langle \mathbb{A}_1(\cdot,x)e_k,B\mathbb{A}_2(\cdot,x')e_j\rangle_{L^2(\R^2_+,\C^2)}\,.
	\end{split}\]
\end{lemma}
\begin{proof}
	\noindent \textbf{Continuity.} Let us first show that the map
    \begin{equation}\label{pim6}
    \R_+^2\times \R^2_+\ni (x,x')\mapsto \mathcal{K}_{kj}(x,x')=\int_{\R^2_+}\langle e_k,\mathbb{A}_1^\dagger(y,x)\big(B\mathbb{A}_2(\cdot,x')e_j\big )(y)\rangle_{\C^2}  \mathrm{d}y\,
    \end{equation}
    is continuous.
    
    Indeed, we have
	\begin{equation}\label{pim3}\begin{split}
		&\langle \mathbb{A}_1(\cdot,x)e_k,B\mathbb{A}_2(\cdot,x')e_j\rangle-\langle \mathbb{A}_1(\cdot,w)e_k,B\mathbb{A}_2(\cdot,w')e_j\rangle\\
		=&\langle \mathbb{A}_1(\cdot,x)e_k,B\mathbb{A}_2(\cdot,x')e_j\rangle-\langle \mathbb{A}_1(\cdot,x)e_k,B\mathbb{A}_2(\cdot,w')e_j\rangle\\
		&+\langle \mathbb{A}_1(\cdot,x)e_k,B\mathbb{A}_2(\cdot,w')e_j\rangle-\langle \mathbb{A}_1(\cdot,w)e_k,B\mathbb{A}_2(\cdot,w')e_j\rangle
	\end{split}
    \end{equation}
	For the first difference we have
	\begin{equation*}\begin{split}|\langle \mathbb{A}_1(\cdot,x)e_k,B\mathbb{A}_2(\cdot,x')e_j\rangle-\langle \mathbb{A}_1(\cdot,x)e_k,B\mathbb{A}_2(\cdot,w')e_j\rangle|\\
		\leq \|\mathbb{A}_1(\cdot,x)e_k\|
        _{L^2}\,\| B\|_{\mathcal{B}(L^2)}\, \| \mathbb{A}_2(,\cdot,x')e_j-\mathbb{A}_2(\cdot,w')e_j\|_{L^2}\,.\end{split}
        \end{equation*}
	This term goes to zero when $w'\to x'$ due to the assumption on $\mathbb{A}_2(\cdot,x')$. The second term in \eqref{pim3} can be treated in a similar way by using that $B^*$ is bounded and the assumption on $\mathbb{A}_1(\cdot,x)$. 
    
	\noindent \textbf{Proving the kernel formula.} Let $f\in\mathcal{C}^\infty_0(\R^2_+,\C^2)$. 
For all $\delta>0$ and all $\ell\in\mathbb{Z}^2$, we define $\mathrm{Box}_{\ell\, \delta}=\ell \, \delta +[0,\delta]^2$. We also write $|B_f(\ell\, \delta)|:={\rm area}\big (\mathrm{Box}_{\ell\, \delta}\cap\mathrm{supp} f\big )$. 
    
    First, we would like to prove that  in the topology of $L^2(\R^2_+,\mathbb{C}^{2})$ we have
	\begin{equation}\label{pim5}A_jf(x)=\lim_{\delta\to 0} \sum_{\ell\in\mathbb{Z}^2} |B_f(\ell\, \delta)|\, \mathbb{A}_j(x,\ell\, \delta)f(\ell\, \delta)\,.
    \end{equation}
In order to do so, we write
	\[\begin{split}A_j f(x)&=\int_{\R^2_+}\mathbb{A}_j(x,y)f(y)\mathrm{d}y=\sum_{\ell\in\mathbb{Z}^2}\int_{\mathrm{Box}_{\ell\, \delta}\cap\mathrm{supp} f}\mathbb{A}_j(x,y)f(y)\mathrm{d}y\\
		&=\sum_{\ell\in\mathbb{Z}^2}\int_ {\mathrm{Box}_{\ell\delta}\cap\mathrm{supp}f}\mathbb{A}_j(x,y)f(\ell\, \delta)\mathrm{d}y+R_\delta(x)\\
		&=\sum_{\ell\in\mathbb{Z}^2}\mathbb{A}_j(x,\ell\delta)f(\ell\, \delta)\, |B_f(\ell\, \delta)|+\tilde R_\delta(x)+R_\delta(x)\, ,	
	\end{split}\]
	where
\[R_\delta(x)=\sum_{\ell\in\mathbb{Z}^2}\int_{\mathrm{Box}_{\ell\, \delta}\cap\mathrm{supp} f}\mathbb{A}_j(x,y)\big (f(y)-f(\ell\, \delta)\big )\mathrm{d}y\]
	and
	\[\tilde R_\delta(x)=\sum_{\ell\in\mathbb{Z}^2}\int_{\mathrm{Box}_{\ell \delta}\cap\mathrm{supp}f}\big (\mathbb{A}_j(x,y)-\mathbb{A}_j(x,\ell\, \delta)\big )f(\ell\, \delta)\mathrm{d}y\, .\]
	Let us now estimate $\|R_\delta\|_{L^2}$. Denoting the matrix operator norm by $\|\cdot\|_{\C^{2\times 2}}$ and 
    $$M_j:=\sup_{x'\in \R^2_+}\, \Big (\int_{\R^2_+}\|\mathbb{A}_j(x,x')\|^2_{\C^{2\times 2}}\, \mathrm{d}x\Big )^{1/2}$$
    we have
	\[\begin{split}\|R_\delta\|&\leq \sum_{\ell\in \mathbb{Z}^2}\int_{\mathrm{Box}_{\ell \, \delta}\cap\mathrm{supp} f}\Big (\int_{\R^2_+}\|\mathbb{A}_j(x,y)\|^2_{\C^{2\times 2}}\, \mathrm{d}x\Big )^{1/2}\, |f(y)-f(\ell\, \delta)|\mathrm{d}y\\
		&\leq M_j \sum_{\ell\in \mathbb{Z}^2}\int_{\mathrm{Box}_{\ell\, \delta}\cap\mathrm{supp} f}|f(y)-f(\ell\, \delta)|\mathrm{d}y\\
		&\leq M_j C \delta \sum_{\ell\in \mathbb{Z}^2}\int_{\mathrm{Box}_{\ell\, \delta}\cap\mathrm{supp} f}\mathrm{d}y\leq  M_j \,  C \delta\, {\rm area}({\rm supp}(f)).
	\end{split}
	\]
	Moreover,
	\[\|\tilde R_\delta\|\leq\sum_{\ell\in \mathbb{Z}^2}|f(\ell\, \delta)|\int_{\mathrm{Box}_{\ell\, \delta}\cap\mathrm{supp} f}\Big (\int_{\R^2_+}\|\mathbb{A}_j(x,y)-\mathbb{A}_j(x,\ell\, \delta)\|_{\C^{2\times 2}}^2 \, \mathrm{d} x\Big )^{1/2}\mathrm{d}y\, .\]
	The uniform continuity of $y\mapsto\mathbb{A}_j(\cdot,y)$ on $\mathrm{supp} f$ implies that $\|\tilde R_\delta\|\to 0$ with $\delta$, which finishes the proof of \eqref{pim5}.

    Now, for $f,g\in\mathcal{C}^\infty_0(\R^2_+,\C^2)$ we have $\langle f,A_1^*BA_2g\rangle=\langle A_1 f,BA_2 g\rangle$, and using the uniform continuity on compacts of the map in \eqref{pim6} we have 
	\[\begin{split}\langle A_1 f,BA_2 g\rangle&=\lim_{\delta\to 0}\sum_{\ell,\ell'\in \mathbb{Z}^2}|B_f(\ell \, \delta)|\, |B_g(\ell' \, \delta)|\, \langle\mathbb{A}_1(\cdot,\ell\, \delta)f(\ell\, \delta),B\mathbb{A}_2(\cdot,\ell'\, \delta)g(\ell'\, \delta)\rangle_{L^2(\R^2_+,\mathbb{C}^2)}\\
    &=\lim_{\delta\to 0}\sum_{j,k=1}^2\sum_{\ell,\ell'\in \mathbb{Z}^2}|B_f(\ell \, \delta)|\, |B_g(\ell' \, \delta)|\,  \overline{f_j(\ell\, \delta)}\,  \mathcal{K}_{jk}(\ell\, \delta,\ell'\, \delta)g_k(\ell'\, \delta)\\
		&=\sum_{j,k=1}^2\int_{\mathbb{R}^2_+}\int_{\R^2_+}  \overline{f_j(x)}\,  \mathcal{K}_{jk}(x,x')g_k(x')\mathrm{d}x' \, \mathrm{d}x\,,\end{split}\]
	hence the matrix $\big \{\mathcal{K}_{jk}(x,x')\big \}_{1\leq j,k\leq 2}$
    is the integral kernel of $A_1^*BA_2$. 
\end{proof}

\begin{lemma}[Combes--Thomas estimate]\label{lem.CT}
Let $b\in\R$ and $W\in\mathcal{C}_b(\R_+^2,\C^{2\times 2})$ be a two by two Hermitian matrix-valued function. Let $C_1\in(0,1)$. For all $z=u+iv$ with $v\in \R\setminus\{0\}$ and all $x_0\in\R^2_+
	$, we have
	\[\|e^{C_1|v|\langle\cdot-x_0\rangle}(\mathscr{D}_{b,W}-z)^{-1}e^{-C_1|v|\langle\cdot-x_0\rangle}\|\leq (1-C_1)^{-1}|v|^{-1}\,.\]
\end{lemma}
\begin{proof}
In order to simplify notations we set $\mathscr{D}\equiv \mathscr{D}_{b,W}$. 
	Let $C_1>0$. On the domain of $\mathscr{D}$ we have
	\[\begin{split}\mathscr{D}_{\mathrm{conj}}-z& \equiv e^{C_1|v|\langle\cdot-x_0\rangle}(\mathscr{D}-z)e^{-C_1|v|\langle\cdot-x_0\rangle}\\
    &=(\mathscr{D}-z)+iC_1|v|\sigma\cdot\nabla\langle\cdot-x_0\rangle =\left(1+iC_1|v|[\sigma\cdot\nabla\langle\cdot-x_0\rangle](\mathscr{D}-z)^{-1}\right)(\mathscr{D}-z)\,.
	\end{split}\]
	Then, we observe that $\|\sigma\cdot\nabla\langle\cdot-x_0\rangle\|_{\C^{2\times 2}}=\frac{|\cdot-x_0|}{\langle\cdot-x_0\rangle}\leq 1$. Moreover, $\|(\mathscr{D}-z)^{-1}\|\leq |v|^{-1}$.
	Hence, we have
	\[
		\|T\|\leq C_1\,,\quad T(z)=iC_1|v|[\sigma\cdot\nabla\langle\cdot-x_0\rangle](\mathscr{D}-z)^{-1}\,.
	\]
	For $C_1\in(0,1)$, the operator $(1+T)$ is invertible and $\|(1+T)^{-1}\|\leq (1-C_1)^{-1}$. We conclude that the operator 	$\mathscr{D}_{\mathrm{conj}}-z$ is bijective from $\mathrm{Dom}(\mathscr{D})$ to $L^2(\R^2_+,\C^2)$, 
	\[(\mathscr{D}_{\mathrm{conj}}-z)^{-1}=(\mathscr{D}-z)^{-1}(1+T(z))^{-1}\]
	and
	\[\|(\mathscr{D}_{\mathrm{conj}}-z)^{-1}\|\leq \frac{1}{(1-C_1)|v|}\,.\]
\end{proof}

\begin{lemma}\label{lemma-pim}
Let $A_1$ and $A_2$ be two integral operators with jointly continuous integral kernels which fulfill the conditions of Lemma \ref{lem.Simon}. Assume further that there exists $\mu>0$ and $C>0$ such that their integral kernels satisfy $|\mathbb{A}_j(x,x')|\leq C\, e^{-\mu|x-x'|}$ for all $x,x'\in \R^2_+$. 

Then the operator $A_1^* \, (\mathscr{D}_{b,W}-z)^{-1}\, A_2$ with $z=u+iv\in \C$ with $v\neq 0$ has a jointly continuous integral kernel. Moreover, there exist $d_1,d_2>0$ such that for all $z=u+iv$ with $0<|v|\leq 1$ we have
$$\Big |\big (A_1^*\, (\mathscr{D}_{b,W}-z)^{-1}\, A_2\big )(x,x')\Big |\leq d_2\, |v|^{-1}\,  e^{-d_1\, |v|\, |x-x'|}\, .$$
\begin{proof}
The integral kernel of $A_1^*(\mathscr{D}_{b,W}-z)^{-1}\, A_2$ equals  
$$\mathcal{K}_{jk}(x,x')=\langle \mathbb{A}_1(\cdot,x)e_j,\, (\mathscr{D}_{b,W}-z)^{-1}\,\mathbb{A}_2(\cdot,x')e_k\rangle_{L^2(\R^2_+,\C^2)}.$$
Let us choose $C_1\in (0,1)$ from Lemma \ref{lem.CT} to be less than $\mu$. We put $d_1:=C_1<\mu$. Then we may write $\mathcal{K}_{jk}(x,x')$ as
$$\langle e^{-d_1 \, |v|\, |\cdot -x'|}\, \mathbb{A}_1(\cdot,x)e_j,\, \Big (e^{d_1 \, |v|\, |\cdot -x'|}\, (\mathscr{D}_{b,W}-z)^{-1}\, e^{-d_1 \, |v|\, |\cdot -x'|}\Big )\,  e^{d_1 \,|v|\,  |\cdot -x'|}\mathbb{A}_2(\cdot,x')e_k\rangle_{L^2(\R^2_+,\C^2)}.$$
Applying the Cauchy--Schwarz inequality and the operator norm estimate in Lemma \ref{lem.CT} we obtain that there exists $C>0$ such that  
$$|\mathcal{K}_{jk}(x,x')|\leq C \, |v|^{-1}\, \|e^{-d_1 \, |v|\, |\cdot -x'|}\, \mathbb{A}_1(\cdot,x)e_j\|_{L^2(\R^2_+,\C^2)}, $$
which together with the triangle inequality $|x-x'|\leq |\cdot -x|+|\cdot -x'|$ leads to
$$e^{d_1\,|v|\,  |x-x'|}\, |\mathcal{K}_{jk}(x,x')|\leq C \, |v|^{-1}\,\|e^{d_1 \, |v|\,   |\cdot -x|}\, \mathbb{A}_1(\cdot,x)e_j\|_{L^2(\R^2_+,\C^2)}\leq \frac{d_2}{|v|}.$$
\end{proof}
\begin{remark}\label{remark-pim}
In a certain particular case, namely when the exponent $\mu=\mu_0\, |v|$ with some $\mu_0>0$, we may allow $|v|$ to be larger than one. Indeed, from the proof of Lemma \ref{lemma-pim} we see that the only condition to be satisfied is $d_1\, |v|<\mu=\mu_0\,|v|$, which holds if  $d_1<\mu_0$. 
\end{remark}

\end{lemma}

\bibliographystyle{abbrv}
\bibliography{cinq}

@article {MR3950662,
    AUTHOR = {Stockmeyer, Edgardo and Vugalter, Semjon},
     TITLE = {Infinite mass boundary conditions for {D}irac operators},
   JOURNAL = {J. Spectr. Theory},
  FJOURNAL = {Journal of Spectral Theory},
    VOLUME = {9},
      YEAR = {2019},
    NUMBER = {2},
     PAGES = {569--600},
      ISSN = {1664-039X,1664-0403},
   MRCLASS = {81Q10 (35Q41 46N50)},
  MRNUMBER = {3950662},
       DOI = {10.4171/JST/256},
       URL = {https://doi.org/10.4171/JST/256},
}

@article {MR670130,
    AUTHOR = {Simon, Barry},
     TITLE = {Schr\"odinger semigroups},
   JOURNAL = {Bull. Amer. Math. Soc. (N.S.)},
  FJOURNAL = {American Mathematical Society. Bulletin. New Series},
    VOLUME = {7},
      YEAR = {1982},
    NUMBER = {3},
     PAGES = {447--526},
      ISSN = {0273-0979,1088-9485},
   MRCLASS = {81-02 (35-02 35P05 47D05 47F05 81C10)},
  MRNUMBER = {670130},
MRREVIEWER = {Ren\'e\ Carmona},
       DOI = {10.1090/S0273-0979-1982-15041-8},
       URL = {https://doi.org/10.1090/S0273-0979-1982-15041-8},
}

@book{vladimirov1971equations,
  title={Equations of Mathematical Physics},
  author={Vladimirov, V.S.},
  isbn={9780824717131},
  lccn={76134446},
  series={Monographs and textbooks in pure and applied mathematics},
  url={https://books.google.it/books?id=oAfvAAAAIAAJ},
  year={1971},
  publisher={M. Dekker}
}

@article{barbaroux2019resolvent,
	author = {Barbaroux, Jean-Marie and Cornean, Horia and Le Treust, Lo{\"\i}c and Stockmeyer, Edgardo},
	date = {2019/06/01},
	doi = {10.1007/s00023-019-00787-2},
	id = {Barbaroux2019},
	isbn = {1424-0661},
	journal = {Annales Henri Poincar{\'e}},
	number = {6},
	pages = {1877--1891},
	title = {Resolvent Convergence to {D}irac Operators on Planar Domains},
	url = {https://doi.org/10.1007/s00023-019-00787-2},
	volume = {20},
	year = {2019}}

@article {MR4310815,
    AUTHOR = {Cornean, Horia D. and Monaco, Domenico and Moscolari, Massimo},
     TITLE = {Beyond {D}iophantine {W}annier diagrams: gap labelling for
              {B}loch--{L}andau {H}amiltonians},
   JOURNAL = {J. Eur. Math. Soc. (JEMS)},
  FJOURNAL = {Journal of the European Mathematical Society (JEMS)},
    VOLUME = {23},
      YEAR = {2021},
    NUMBER = {11},
     PAGES = {3679--3705},
      ISSN = {1435-9855,1435-9863},
   MRCLASS = {81Q10 (81Q15 81Q30 81Q70 81R10)},
  MRNUMBER = {4310815},
MRREVIEWER = {Lawrence\ E.\ Thomas},
       DOI = {10.4171/jems/1079},
       URL = {https://doi.org/10.4171/jems/1079},
}

@inproceedings{cornean2024orbital,
  title={From orbital magnetism to bulk-edge correspondence},
  author={Cornean, Horia D and Moscolari, Massimo and Teufel, Stefan},
  booktitle={Annales Henri Poincar{\'e}},
  pages={1--55},
  year={2024},
  organization={Springer}
}

@article{barbaroux:hal-02889558,
  title={The {D}irac bag model in strong magnetic fields},
  author={Barbaroux, Jean-Marie and Le Treust, Lo{\"\i}c and Raymond, Nicolas and Stockmeyer, Edgardo},
  journal={Pure and Applied Analysis},
  volume={5},
  number={3},
  pages={643--727},
  year={2023},
  publisher={Mathematical Sciences Publishers}
}

@article{davies1995functional,
  title={The functional calculus},
  author={Davies, Edward B},
  journal={Journal of the London Mathematical Society},
  volume={52},
  number={1},
  pages={166--176},
  year={1995},
  publisher={Citeseer}
}

@article {MR4769230,
    AUTHOR = {Barbaroux, J.-M. and Cornean, H. D. and Le Treust, L. and
              Raymond, N. and Stockmeyer, E.},
     TITLE = {Magnetic {D}irac systems: violation of bulk-edge
              correspondence in the zigzag limit},
   JOURNAL = {Lett. Math. Phys.},
  FJOURNAL = {Letters in Mathematical Physics},
    VOLUME = {114},
      YEAR = {2024},
    NUMBER = {4},
     PAGES = {Paper No. 93, 27},
      ISSN = {0377-9017,1573-0530},
   MRCLASS = {81Q10 (35P30 35Q40 81T35)},
  MRNUMBER = {4769230},
MRREVIEWER = {C\'esar\ R.\ de Oliveira},
       DOI = {10.1007/s11005-024-01839-3},
       URL = {https://doi.org/10.1007/s11005-024-01839-3},
}

@article{Cornean_2023,
	doi = {10.1063/5.0119022},
  
	url = {https://doi.org/10.1063%2F5.0119022},
  
	year = 2023,
	month = {{F}eb},
  
	publisher = {{AIP} Publishing},
  
	volume = {64},
  
	number = {2},
  
	pages = {021902},
  
	author = {H. D. Cornean and M. Moscolari and K. S. S{\o}rensen},
  
	title = {Bulk-edge correspondence for unbounded {D}irac{\textendash}{L}andau operators},
  
	journal = {Journal of Mathematical Physics}
}
    \end{document}